\documentclass[11pt]{article}
\usepackage{hyperref}
\pdfoutput=1

\usepackage[english]{babel}
\usepackage{amsmath}
\usepackage{amssymb}
\usepackage{graphicx}
\usepackage{natbib}
\usepackage[nolists,tablesfirst, noheads]{endfloat}
\usepackage{appendix, bbm, color}
\usepackage{fullpage, multirow}
\usepackage{authblk}

\providecommand{\keywords}[1]{\textbf{\textit{Keywords:}} #1}
\def\bnu{\mbox{\boldmath $\nu$}}
\def\btau{\mbox{\boldmath $\tau$}}
\def\bgamma{\mbox{\boldmath $\gamma$}}
\def\balpha{\mbox{\boldmath $\alpha$}}
\def\blambda{\mbox{\boldmath $\lambda$}}

\def\bphi{\mbox{\boldmath $\phi$}}

\def\btheta{\mbox{\boldmath $\theta$}}
\def\bGamma{\mbox{\boldmath $\Gamma$}}
\def\bLambda{\mbox{\boldmath $\Lambda$}}
\def\bTheta{\mbox{\boldmath $\Theta$}}
\def\balpha{\mbox{\boldmath $\alpha$}}

\def\bbeta{\mbox{\boldmath $\beeta$}}
\def\bvarepsilon{\mbox{\boldmath $\varepsilon$}}

\def\beeta{\mbox{\boldmath $\eta$}}
\def\bSigma{\mbox{\boldmath $\Sigma$}}

\def\bOmega{\mbox{\boldmath $\Omega$}}

\def\bSigma{\mbox{\boldmath $\Sigma$}}
\def\bZero{\mbox{\boldmath $0$}}
\def\bI{\mbox{\boldmath $I$}}
\def\bD{\mbox{\boldmath $D$}}

\def\bvarepsilon{\mbox{\boldmath $\varepsilon$}}
\def\bepsilon{\mbox{\boldmath $\epsilon$}}
\def\bc{\mbox{\boldmath $c$}}
\def\bgamma{\mbox{\boldmath $\gamma$}}
\def\bGamma{\mbox{\boldmath $\Gamma$}}
\def\bvarpi{\mbox{\boldmath $\varpi$}}

\def\bB{\mathbf{B}}
\def\bW{\mathbf{W}}
\def\bZ{\mathbf{Z}}
\def\bC{\mathbf{C}}
\def\bbb{\mathbf{b}}
\def\bzr{\mathbf{0}}
\def\by{\mathbf{y}}
\def\bY{\mathbf{Y}}
\def\bZ{\mathbf{Z}}
\def\bW{\mathbf{W}}
\def\bM{\mathbf{M}}
\def\bV{\mathbf{V}}

\def \diag{\mathop{\rm diag}}

\def \al{\text{\rm -- }}

\begin{document}
\title{High-dimensional Bayesian Fourier Analysis For Detecting Circadian Gene Expressions}
\author[1]{Silvia Montagna\thanks{simontag@unimore.it}}
\author[2]{Irina Irincheeva}
\author[3]{Surya T. Tokdar}
\affil[1]{Dipartimento di Economia ``Marco Biagi'', Universit\`a di Modena e Reggio Emilia, via Berengario 51, 41122 Modena}
\affil[2]{Clinical Trials Unit (CTU Bern), University of Bern, Bern 3012, Switzerland}
\affil[3]{Department of Statistical Science, Duke University, Box 90251, Durham NC 27708, USA}

\renewcommand\Authands{ and }

\maketitle

\begin{abstract}
In genomic applications, there is often interest in identifying genes whose time-course expression trajectories exhibit periodic oscillations with a period of approximately 24 hours. Such genes are usually referred to as ``circadian'', and their identification is a crucial step toward discovering physiological processes that are clock-controlled. It is natural to expect that the expression of gene $i$ at time $j$ might depend to some degree on the expression of the other genes measured at the same time. However, widely-used rhythmicity detection techniques do not accommodate for the potential dependence across genes. We develop a Bayesian approach for periodicity identification that explicitly takes into account the complex dependence structure across time-course trajectories in gene expressions. We employ a latent factor representation to accommodate dependence, while representing the true trajectories in the Fourier domain allows for inference on period, phase, and amplitude of the signal. Identification of circadian genes is allowed through a carefully chosen variable selection prior on the Fourier basis coefficients. The methodology is applied to a novel mouse liver circadian dataset. Although motivated by time-course gene expression array data, the proposed approach is applicable to the analysis of dependent functional data at broad.
\end{abstract}


\section{Introduction} 
Circadian rhythms are cycles of biological activity based on a 24-hour period which allow organisms to anticipate and adapt to predictable daily oscillations in the environment \citep{Hughes2010}. Circadian rhythms are controlled by the circadian clock, namely a network of mutually interacting proteins that generate transcriptional and translational feedback loops \citep{Fred2013}. The molecular mechanisms underlying the circadian clock have been investigated in many organisms \citep{Wichert2004, Fred2013}. In plants, circadian rhythmicity has been extensively studied in the {\textit{Arabidopsis thaliana}} \citep{Anderson2006, Edwards2006, Liverani2009}. In animals, sleep-wake cycles are circadian-regulated to maximize the availability of food as well as to avoid predation. In humans, blood pressure, hormone production, metabolism and other biological cycles are clock-regulated, and disruptions to the circadian rhythms have been linked to a variety of pathologies \citep{Hughes2010}. For example, the International Agency for Research on Cancer reports that ``shift-work that involves circadian disruption is probably carcinogenic to humans'' \citep{2007IARC}. Consequently, there is a considerable interest in identifying genes that control the timing of many physiological processes. 

The identification of clock-genes is performed through examination of their expression levels or ``transcripts''. With regard to the functioning of a cell, deoxyribonucleic acid (DNA) is first duplicated into messenger ribonucleic acid (mRNA), and the RNA is then used for protein synthesis. To quantify the expression of a specific gene, it is possible to measure the concentration of RNA molecules associated with this gene. By using this principle, microarray analysis allows investigators to measure many hundreds or thousands of transcripts simultaneously, and then statistical methods are needed to detect periodic pathways among a very high number of gene expression profiles.   

Several authors have proposed methods for periodicity identification in biomedical research over the last couple of decades. \cite{Chudova2009} give an excellent review of the main existing techniques, which can be broadly classified as time domain or frequency domain analyses. Time domain methods are pattern-matching techniques: cosine curves of varying periods and phases are fit to each expression profile separately, and the best fit to the experimental data is retained to describe the signal \citep{Straume2004, Hughes2010}. Pattern-matching methods are simple and computationally efficient, but not very effective at finding periodic signals that are not perfectly sinusoidal \citep{Chudova2009}. Frequency domain approaches combine spectral analysis with multiple hypothesis testing \citep{Wichert2004, Olli2005}. Specifically, one obtains the spectrum of a transcript and the hypothesis of significance of the dominant frequency is tested against the null hypothesis of absence of periodic signal. The analysis is carried out probe by probe independently, and the obtained significance values are then corrected for multiplicity. \cite{Chudova2009} remark that frequency domain methods are most effective on long time series. However, this is not a typical feature of circadian studies, which are usually designed to collect data every 2 or 4 hours over two circadian cycles. Therefore, coarse sampling and short periods of data collection are typical features of these studies. Also, the authors claim that existing computational methods are biased toward discovering genes whose transcripts follow sine-wave patterns. \cite{Chudova2009} propose an analysis of variance periodicity detector and its Bayesian extension for the identification of patterns of arbitrary shape. \cite{Costa2013a} propose a methodology based on spectrum resampling (SR). SR is a technique based around an iterative bootstrapping of a smoothed power series to provide a steady-state model which reflects the data. The authors point out that SR is especially designed to be robust to non-sinusoidal and noisy cycles.

As a separate line of research for the analysis of genomic data, several model-based clustering algorithms have been proposed in both the classical and Bayesian framework \citep{Yeung2001, Luan2003, Wakefield2003}. However, clustering algorithms should be customized to reflect the scientific interest. In particular, efforts should concentrate around refining clusters that contain potentially interesting genes while no time should be spent on finding an optimal partition of obviously non-circadian genes. In this regard, \cite{Anderson2006} use the algorithm in \cite{Heard2006} many times on various partitions of the transcripts, with a Fourier basis to extract rhythmically expressed genes. A score is calculated for each partition of the genes and the score determines the clustering.  

\begin{figure}[hb]
  \centering
  \label{figure1}
  \includegraphics[width = 1\textwidth]{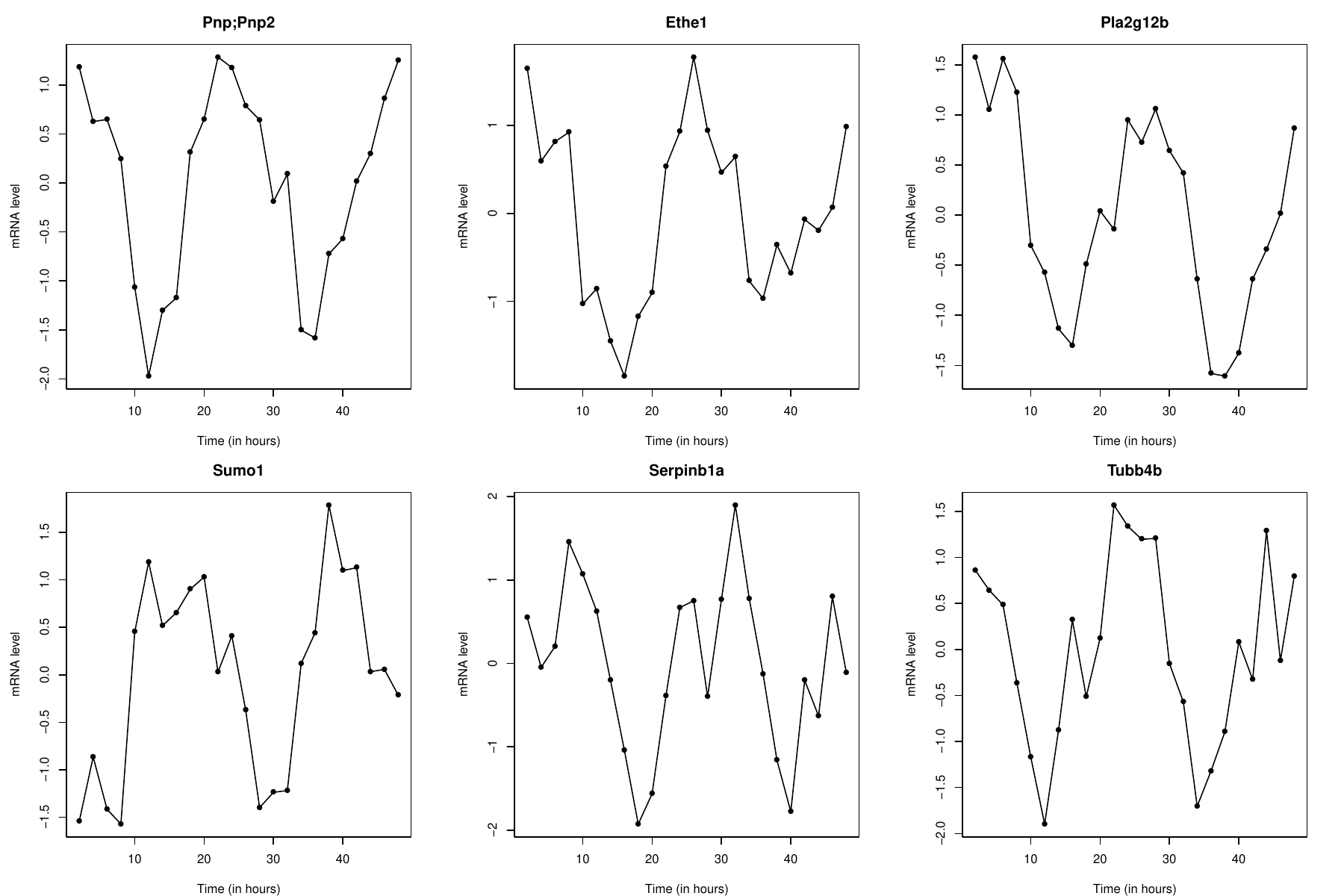}
  \caption{Examples of temporal mRNA expression profiles of rhythmically translated genes as selected by our approach from \cite{Fred2013} microarray dataset. Data are log transformed and normalised following standard procedure.}
\end{figure}

A key assumption in all the approaches above is that of independence across transcripts. The clustering algorithm described in \cite{Anderson2006} assumes dependence at a cluster level only whereas clusters vary independently. Although practical from a computational perspective, the independence assumption is often too strong to be realistic in many applications, including the data set that motivates our work. In particular, \cite{Fred2013} design a microarray experiment to assess whether the circadian clock coordinates mRNA translation in mouse liver. Male mice between 10 and 12 weeks of age were kept under standard animal housing conditions, with free access to food and water and in 12 hours light/12 hours dark cycles. However, animals were fed only at night during 4 days before the experiment to reduce the effects of feeding rhythm. 3 $\mu$g of liver polysomal and total RNAs were extracted independently from two mice sacrificed every 2 hours during 48 hours, and used for the synthesis of biontinylated cytosine RNA according to Affymetrix protocol. The resulting fluorescence signal was analysed with Affymetrix software (refer to \cite{Fred2013} for an extensive description of the experiment). Data are deposited on the Gene Expression Omnibus database under the reference GSE33726. Figure \ref{figure1} shows examples of temporal expression profiles of a subset of genes from the dataset. Being collected on different mice, the measurements across time can reasonably be assumed to be independent. However, it is natural to expect that the expression of gene $i$ at time $j$ might depend to some degree on the expression of the other genes measured at the same time. Therefore, the dependence across expression trajectories should be modelled and explicitly accounted for. 

In this paper, we propose a flexible Bayesian approach that identifies periodic signals in gene expression profiles while accounting for dependence in the functional data. Specifically, we decompose the true, de-noised underlying signal for each transcript as a series expansion of sine and cosine waves to extract rhythmic signals, while we also accommodate for local deviations from these smooth, sinusoidal trajectories. Furthermore, we accommodate conditional dependence across probes at each time point through a latent factor framework. Dimensionality reduction and sparsity are induced through careful modeling of the latent factors as well as the local and Fourier basis coefficients. The proposed approach gives a comprehensive description of cyclic rhythms through posterior summaries of the model parameters. The proposed approached is tested on the mouse liver mRNA dataset of \cite{Fred2013} and on simulation experiments. \\
\indent The rest of the paper is organized as follows. Section~\ref{Motivation} presents our dependent latent factor approach. Section~\ref{priors} discusses the choice of prior distributions for the model parameters, and Section~\ref{Post} outlines the posterior computation. In Section~\ref{lf_simulation} we apply our approach to synthetic datasets. Section~\ref{protein} present the application of our methodology to the mouse liver mRNA dataset. Concluding remarks are presented in Section~\ref{conclusions}. 

\section{Methodology}
\label{Motivation}

\subsection{Overview}
We consider data from a gene expression experiment in the form of a $p \times T$ matrix $\bY = \{y_{ij}\}$, where entry $y_{ij}$ denotes the observed mRNA concentration for gene $i$ at time $t_j$, for $i = 1, \dots, p$, and with $p$ denoting the total number of genes. In circadian microarray studies, data are typically collected over two complete circadian cycles and the sampling rate is usually either two of four hours. In our motivating application \citep{Fred2013}, the sampling rate is two hours, thus $t_j = 0, 2, 4 \dots, 46$ and $T = 24$. Hereafter, we will make explicit reference to the study in \cite{Fred2013}, although the structure applies more generally to any circadian microarray experiment with the appropriate choices of $T$ and sampling rate. We assume that the $y_{ij}$'s are error-prone measurements of an underlying smooth true trajectory
	\begin{equation} y_{ij} = f_i(t_j) + \nu_{ij}. \label{Eq1}
	\end{equation} 
Suppose that the de-trended and centered true signal for gene $i$ at time $t_j$, $f_i(t_j)$, can be decomposed as 

$$f_i(t_j) = \sum_{m = 1}^{q} (\theta_{i,2m-1} b_{2m-1}(t_j)+\theta_{i, 2m} b_{2m}(t_j))=\btheta_{i,m}^\top \bbb_m(t_j)=\btheta_{i,m}^\top \bbb_{m,j},$$ 

where for $m=1, \ldots, q$ we define $\btheta_{i,m}=(\theta_{i,2m-1}, \theta_{i,2m})^\top$ and $\bbb_{m,j}=\left[ b_{2m-1}(t_j),  b_{2m}(t_j)\right]^{\top}$.  
The vector 

$$\bbb_j = [b_1(t_j), b_2(t_j),\dots, b_{2q-1}(t_j), b_{2q}(t_j)]^\top$$

represents a set of $2q$ fixed basis functions evaluated at time $t_j$. One popular basis for a space of periodic functions is the Fourier basis 

$$\bbb(t) = \left[\sin\left(\frac{2\pi}{\omega_1} t \right),
\cos\left(\frac{2\pi}{\omega_1} t \right), \dots,
\sin\left(\frac{2\pi}{\omega_{q}} t \right), \cos\left(\frac{2\pi}{\omega_{q}}
t \right) \right],$$ 

where $\{\omega_m\}_{m = 1}^q$ denotes the periodicity of the signal and $t$ is time represented by a unit-interval increase. The $q$ period lengths $w_m$ are assumed known and fixed. Since there are 24 time points per transcript in the mouse liver ribosomal proteins dataset, we can use up to twelve sine/cosine pairs of harmonics. According to Nyquist-Shannon theorem and common sense, the possible range of periods would be $4, 6, 8, 10, 12,14, 16, 18, 20, 22$ or 24 hours, but biologists would argue to reduce this list to $4, 6, 8, 12$ or 24 hours only. In practice, suitable period lengths can be proposed by inspecting the average periodogram of the probes and choosing the frequencies of the $q$ ordered largest peaks in the spectrum. Here $w_1$ is the shortest period, and $w_2, \dots, w_q$ correspond to longer periods. \\
\indent The term $\nu_{ij}$ in Equation \ref{Eq1} models the deviation between the observed measurement at time $t_j$, $y_{ij}$, and the underlying smooth profile. In the original study \citep{Fred2013}, two mice are sacrificed every two hours and the reported $p$ expression levels at time $j$ are obtained by pooling 3 $\mu$g o total mRNA from each mice. Therefore, the $\nu_{ij}$'s are correlated across probes, $i$. Specifically, each profile at time $j$ may deviate from its own underlying truth because of a ``mouse effect'' which could make, e.g. a certain protein more expressed than the corresponding truth and another protein less expressed at time $j$. To accommodate dependence across probes at time $j$ we adopt a sparse factor model:
\begin{equation}
\bnu_{ j} = \bLambda \beeta_{j} + \bepsilon_{
j},
\label{nu_model}
\end{equation}
 with $\bnu_{ j} = [\nu_{1j}, \dots, \nu_{pj}]^\top$, $\bLambda=\left[
\blambda_1, \ldots,  \blambda_p \right]^\top$ is a $p \times k$ factor loading
matrix with elements $\left\{\lambda_{ih}\right\}_{i=1, \ldots, p; ~h=1,
\ldots, k}$, $\beeta_{j} = (\eta_{1j}, \dots, \eta_{kj})^\top$ is $k \times 1$
vector of latent factors at time $j$ which explains mice-specific deviations
of the expression levels at time $j$ from their corresponding truth (it
explains why proteins at time $j$ may be systematically over- or
under-expressed with respect to the ``truth''), and $\bepsilon_{j}$ is a
residual error. Sparsity here is necessary given the very large $p$, and Section \ref{priors} discusses how sparsity can be achieved through the modeling of $\bLambda$. \\

\noindent The full model for subject $i$ at time $t_j$ is
 \begin{align}
 \label{full_model_scalar}
 y_{ij} &= g_i(t_j)+ \blambda_i^\top\beeta_{j}+\epsilon_{ij}, \quad \text{with}  \quad \epsilon_{ij}\sim N(0, \sigma_i^2), \\ \nonumber
 g_i(t_j) &= f_i(t_j) + \bc_j^\top \bgamma_i = \bbb_j^\top\btheta_i + \bc_j^\top \bgamma_i, 
 \end{align}
 where the first term $\bbb_j^\top\btheta_i=\bbb(t_j)^\top \btheta_i$ captures periodic
oscillations, the second term $\bc_j^\top \bgamma_i = \bc(t_j)^\top \bgamma_i$ captures local deviations from the underlying periodic oscillation (if present), and the third term $
\blambda_i^\top \beeta_{j}$ captures across-proteins dependence (if present). The importance of the $\bc_j^\top \bgamma_i$ component becomes evident in studies where mice are given a stimulus at the beginning of the experiment. The stimulus might produce deviations of the observed expression levels from the true signals, and these deviations shall manifest at different times across proteins and last for a different amount of time, if present at all. After standardizing the time domain to $[0,1]$, suitable choices for $\bc(t_j)^\top$ are Gaussian kernels
\begin{equation}
\label{Gaussian}
c_l(t_j) = \exp\{-\psi\vert\vert t_j - \xi_l \vert\vert^2\}, \qquad l = 1, \dots, \tilde{T}
\end{equation}
with equally spaced kernel location $\xi_1, \dots, \xi_{\tilde{T}}$ and bandwidth parameter $\psi$, or B-splines basis functions. $\btheta_i$ ($\bgamma_i$) is the $2q\times 1$ ($\tilde{T} \times 1$) vector of fixed periodic (local) basis function coefficients for protein $i$. Greater $\tilde{T}$ corresponds to more flexibility in modeling local deviations. 
We follow standard practice in normalizing the data prior to analysis and hence do not include an intercept term in (\ref{full_model_scalar}).
We use $\by_i=(y_{i1}, \ldots, y_{iT})^\top$ to denote the
$i$-th row of $\bY$ (the $i$-th protein observed at times $1, \ldots, T$ );
and $\by^{(j)}=(y_{1j}, \ldots, y_{pj})^\top$ to denote the $j$-th column of
$\bY$ ($p$ proteins observed at the time $j$).  Construction
(\ref{full_model_scalar}) for $y_{ij}$ can now be rewritten in vector notation as
\begin{eqnarray}
 \label{yi_model}
 \by_i &=& \bB\btheta_i + \bC\bgamma_i +\bnu_i= \bB\btheta_i+ \bC\bgamma_i + \beeta\blambda_i
+\bvarepsilon_i, \\
   \nonumber &&\text{where} \\
  \nonumber &&\bB=\begin{bmatrix}
 \bbb_1^\top \\
 \vdots \\
 \bbb_{T}^\top  \\
 \end{bmatrix}\in \Re^{T\times 2q};   ~~  \bC=\begin{bmatrix}
 \bc_1^\top \\
 \vdots \\
 \bc_{T}^\top  \\
 \end{bmatrix}\in \Re^{T\times \tilde{T}}; ~~\blambda_i\in \Re^{k};~~\beeta=\begin{bmatrix}
 {\beeta_{1}}^\top \\
 \vdots \\
  {\beeta_{T}}^\top  \\
 \end{bmatrix}\in \Re^{T\times k}; \\
 \nonumber && \bvarepsilon_i=(\varepsilon_{i1}, \ldots,
\varepsilon_{iT})^\top\sim N_T(\bzr, \sigma_i^2 \bI_T)~\text{with $T\times T$
identity matrix} ~\bI_T; \\ \nonumber
\end{eqnarray}

or
\begin{eqnarray}
 \label{yj_model}
 \by^{(j)} &=& \bTheta\bbb_j+ \bGamma \bc_j + \bnu^{(j)}=\bTheta\bbb_j+  \bGamma \bc_j + \bLambda
\beeta_{j}+\bvarepsilon^{(j)},\\
 \nonumber &&\text{where} \\
   \nonumber        &&\bTheta=\begin{bmatrix} \btheta_1^\top \\ \vdots \\
\btheta_p^\top  \\ \end{bmatrix}\in \Re^{p\times 2q}; 
~~ \bGamma=\begin{bmatrix} \bgamma_1^\top \\ \vdots \\
\bgamma_p^\top  \\ \end{bmatrix}\in \Re^{p\times \tilde{T}};
~~\bLambda
=\begin{bmatrix} \blambda_1^\top \\ \vdots \\ \blambda_p^\top  \\
   \end{bmatrix}\in \Re^{p\times k};\\
   \nonumber && \bvarepsilon^{(j)}=(\varepsilon_{1j}, \ldots,
\varepsilon_{pj})^\top\sim N_p(\bzr, \bSigma), ~ \bSigma=\diag\{\sigma_1^2,
\ldots, \sigma_p^2\}
\end{eqnarray}

The latent factors $\beeta_j$ have a natural interpretation as (mice-specific) unobserved traits of the two mice sacrificed at time $j$ that explain the dependence structure across proteins at time $t_j$. Hereafter we follow standard practice and assign a normal prior to the latent factors at time $t_j$, $\beeta_j \sim N(\bzr, \bI_k)$. Proteins are assumed to be independent given the latent factors, and dependence among proteins is induced by marginalizing over the distribution of the factors, so marginally $\by^{(j)} \sim N(\bTheta\bbb_j+ \bGamma \bc_j, \bLambda\bLambda^\top + \bSigma)$. In practical applications involving moderate to large $p$, the number of factors $k$ is typically much smaller than $p$, thus inducing a sparse characterization of the unknown covariance matrix $\bLambda\bLambda^\top + \bSigma$. \\
\indent The next Section discusses suitable prior choices for the model parameters in Equations (\ref{yi_model})-(\ref{yj_model}) and examines how these choices translate into the ability to perform period identification.

\section{Prior elicitation}
\label{priors}

With regard to the modeling of the basis coefficients $\{\btheta_i, \bgamma_i\}_{i = 1}^p$, we need a technical device to induce sparsity / parsimony, hence avoid over-fitting, whilst retaining an easy interpretation of the method. The latter is particularly crucial for the modeling of $\btheta_i$ since inference on this set of parameters is the primary interest of our work. Although different formulations are possible, we adopt the latent threshold model (LTM) of \cite{Nakajima2013}. The LTM is a direct extension of standard Bayesian variable selection which assigns non-zero prior probabilities to zero values of regression parameters, and continuous priors centered at zero otherwise. \\
\indent We begin introducing the LTM for the elements of $\bgamma_i$. Denote with $\gamma_{il}$ the $l$th component of the $\tilde{T} \times 1$ vector of local basis coefficients $\bgamma_i$. The model assumes 
       \begin{equation}
	\label{prior_on_gamma}
	\gamma_{i,l} = \tilde{\gamma}_{i,l}\mathbbm{1}(\vert \tilde{\gamma}_{i,l} \vert \geq \varpi^*_{i,l}),
	\end{equation}
where $\varpi^*_{i,l} \geq 0$ is a latent threshold and $\mathbbm{1}(\cdot)$ denotes the indicator function. Equation (\ref{prior_on_gamma}) embodies sparsity/shrinkage and parameter reduction when necessary, with the $l$th local basis coefficient shrunk to zero when it falls below a threshold. If the true smooth profile for protein $i$ is given by the oscillatory behavior measured by $\bB\btheta_i$ with no time localized deviations, then each component of vector $\tilde{\bgamma}_i = \{\tilde{\gamma}_{i,l}\}_{l = 1}^{\tilde{T}}$ is expected to be uniquely shrunk to zero. Non-zero components allow for time-localized deviations. The vector $\tilde{\bgamma}_{i}$ is modeled as 
	\begin{equation}
	\label{gamma_tilde_model}
	\tilde{\bgamma}_{i} = \bZ \blambda_i + \balpha_i^\gamma \qquad \text{and} \qquad \balpha_i^\gamma \sim N_{\tilde{T}}(\bzr, \bI),
	\end{equation}
where $\bZ$ is a $\tilde{T}\times k$ matrix and $\blambda_i$ is the vector of factor loadings for protein $i$ as in (\ref{full_model_scalar}). We assign a (multivariate) standard normal prior to the rows of $\bZ$, $\bZ_j^\top \sim N_k(\bzr, \bI)$, $j = 1, \dots, \tilde{T}$.\\
\indent We adopt the same variable selection prior for the periodic basis coefficients. Denote with $\btheta_{im} = \{\theta_{i,2m-1}, \theta_{i, 2m}\}^\top$ the vector of $2m-1$th and $2m$th components of $\btheta_i$, $m = 1, \dots, q$. Thus, $\theta_{i, 2m-1}$ is the coefficient of the $2m-1$th sine basis and $\theta_{i, 2m}$ is the coefficient of the $2m$th cosine basis, both harmonics of period $w_m$. To enhance a correct interpretation of periodicity, we need to switch off $\theta_{i, 2m-1}$ and $\theta_{i, 2m}$ jointly provided that shrinkage is supported by the data. Therefore, we assume:
	\begin{equation}
	\label{prior_on_theta}
	\btheta_{i,m} = \tilde{\btheta}_{i,m}\mathbbm{1}(\vert \vert \tilde{\btheta}_{i,m} \vert \vert \geq \varpi_{i,m}),
	\end{equation}
where $\varpi_{i,m}$ is a latent threshold. The idea behind (\ref{prior_on_theta}) is that the value of the $w_m$-periodic basis coefficients is shrunk to zero when their norm falls below a $m$th- (and protein-) specific threshold.  Suppose, for example, that the set of probable periods is $\{4, 6, 8,
12,  24\}$ hours, i.e. $q=5$. If the $i$th time series $\by_i=(y_{i1}, \ldots,
y_{iT})^\top$ is generated with a true signal $f(t)=A\sin\left(\frac{2\pi}{12}t+\varphi\right)$, which is a 12 hours periodic function with phase
$\varphi$ hours and amplitude $A$, then column $i$ of $\bTheta$ is
expected to contain only two non-zero elements, $\theta_{i,7}$ and
$\theta_{i,8}$, for representing sum of 12-hour periodic basis functions
$b_{7}=\sin\left( \frac{2\pi}{12}t\right), b_{8}=\cos\left(
\frac{2\pi}{12}t\right)$. There are two possibilities of parametrization for
$\theta_{i,7}$ and $\theta_{i,8}$: first $\theta_{i,7}=A \sin\left(
\varphi\right)$, $\theta_{i,8}=A \cos \left( \varphi\right)$; and second
$\theta_{i,7}=A \cos\left( \varphi\right)$, $\theta_{i,8}=A \sin \left(
\varphi\right)$. Clearly, this lack of identifiability in $\btheta_{i,4}$ does
not affect in\-fe\-ren\-ce for phase and amplitude. \\
\indent Further, we assume $\tilde{\btheta}_i = \{\tilde{\btheta}_{i,m}\}_{m = 1}^q$ is modeled as 
	\begin{equation}
	\label{theta_tilde_model_regression}
	\tilde{\btheta}_{i} = \bW \blambda_i + \balpha_i^\theta \qquad \text{and} \qquad \balpha_i^\theta \sim N_{2q}(\bzr, \bI),
	\end{equation}
where $\bW$ is a $2q \times k$ matrix and $\blambda_i$ is the vector of factor loadings for protein $i$. Similar to the structure on $\bZ$, we assume $\bW_j^\top \sim N_k(\bzr, \bI)$, $j = 1, \dots, 2q$. This simple structure on $\bgamma_i$ and $\btheta_{i,m}$ in (\ref{prior_on_gamma})-(\ref{prior_on_theta}) allows to flexibly take into account the dependence among parameters $\bgamma_i$, $\btheta_{i,m}$ and $\lambda_i$. \\
To continue, we adopt a multiplicative gamma process shrinkage prior (MGPSP) on the loadings
\begin{eqnarray}\label{mgpsp}
\nonumber  \lambda_{ih} \vert \phi_{ih}, \tau_h &\sim&
\text{N}(0,\phi_{ih}^{-1}\tau_h^{-1}), \quad \phi_{ih} \sim
\text{Ga}\left(\frac{\rho}{2}, \frac{\rho}{2}\right), \quad \tau_h = \prod_{l =
1}^h\zeta_h \\
\nonumber  \zeta_1 &\sim & \text{Ga}(a_1, 1), \quad \zeta_l \sim
\text{Ga}(a_2, 1), \quad l \geq 2, \quad i = 1, \dots, p,
  \end{eqnarray}
 with $h = 1, \dots, k,$ the number of latent factors. $\tau_h$ is a global
shrinkage parameter for the $h$-th column, and $\phi_{ih}$ is a local
shrinkage parameter for the elements in the $h$-th column. In matrix notation,
row $i$ of $\bLambda$ has prior
     \begin{equation}
    \label{prior_lambda_i}
    \blambda_i^\top\vert \{\phi_{ih}\}_{h = 1}^k, \{\tau_h\}_{h = 1}^k \sim
N_k(\bZero, \bD_i),
    \end{equation}

  with $\bD_i = \text{diag}(\phi_{i1}^{-1}\tau_1^{-1}, \dots, \phi_{ik}^{-1}\tau_k^{-1})$. This prior favors more shrinkage as the column index of $\bLambda$ increases, thus avoids factor splitting by concentrating more and more shrunk loadings in the last columns of $\bLambda$. The MGPSP was originally proposed by \cite{Anirban2011} for sparse modeling of high-dimensional covariance matrices. The authors embedded the MGPSP into an adaptive Gibbs sampler which allowed for block update of the rows of the $\bLambda$ while accounting for an adaptive choice of the number of factors, $k$. The main idea consisted of monitoring the columns $\bLambda$ whose loadings were all within some pre-specified neighborhood of zero. If the number of such columns dropped to zero, one extra column was added to $\bLambda$ and otherwise the redundant columns were discarded. We adopt here the same adaptive block Gibbs sampler to retain only important factors and reduce computing time as a side. Additional details on this prior and the adaptive algorithm can be found in \cite{Anirban2011}. \\
 \indent To conclude the model formulation, we need to specify prior distributions on the latent thre\-shold parameters. The straightforward extension of \cite{Nakajima2013} to our scenario leads to a dependent prior for $\varpi_{i,m}$ (and $\varpi_{i,l}^*$) of the type $\varpi_{i,m} \sim \text{Unif}(0, U_{i,m})$ for $i =1 ,\dots, p$ and $m = 1, \dots, q$, where the upper bound of the uniform prior is function (thus dependent) of other model parameters. \cite{Nakajima2013} give a thorough discussion on the choice of the upper bound $U_{i,m}$ and its impact on the sparsity structure of the model.  We recognize, however, that a dependent prior on the latent thresholds would lead to unnecessary complications in the posterior update of some model parameters whitin our construction. Therefore, we opt for independent priors on the latent thresholds
	\begin{eqnarray}\label{prior_latent_thresholds}
	\varpi_{i,m} &\sim& \text{Unif}(0, K_\theta), \quad i = 1, \dots, p, \text{ and } m = 1, \dots, m\\ 
	\varpi^*_{i,l} &\sim& \text{Unif}(0, K_\gamma), \quad i = 1, \dots, p, \text{ and } l = 1, \dots, \tilde{T} \\ 
	K_\theta &\sim & \text{Pareto}(a_\theta, b_\theta), \label{Kthetaprior}\\
	K_\gamma &\sim & \text{Pareto}(a_\gamma, b_\gamma). \label{Kgammaprior}
	\end{eqnarray}
$K_\theta$ and $K_\gamma$ are fundamental sparsity parameters shared across subjects. Smaller or larger degrees of expected sparsity might be needed depending on the context, thus these parameters need to be inferred from the data. In general, the smaller these parameters are estimated to be the less sparse the model becomes. Clearly, there is no inherent interest in direct inference on the thresholds themselves; the interest is their roles as defining the ability to shrink parameters when the data support sparsity. Correspondingly, there is no interest in the underlying values of the latent $\tilde{\bgamma_i}$ and $\tilde{\btheta}_i$ when below threshold. \\
\indent In addition to sparsity, $K_\theta$ and $K_\gamma$ play an important role in controlling for multiplicity adjustments. When analyzing microarray expression data, tens of thousands of genes are estimated simultaneously, so the problem of multiple testing must be considered. \citep{Muller2006} remark that posterior inference adjusts for multiplicities, and no further adjustment is required, provided that the probability model includes a positive prior probability of non-periodic expression for each protein $i$, and a hyperparameter that defines the prior probability mass for non-periodic expression. In our context, the two conditions are controlled and satisfied  by treating $K_\theta$ and $K_\gamma$ as model parameters with hyperpriors as in (\ref{Kthetaprior})-(\ref{Kgammaprior}).  A discussion on the role and specification of hyperparameters $a_\theta, b_\theta, a_\gamma, b_\gamma$ is deferred to the next Section. \\

\subsection{Prior sparsity probabilities}
\begin{figure}[h!]
 \begin{center}
 \includegraphics[width=1\textwidth]{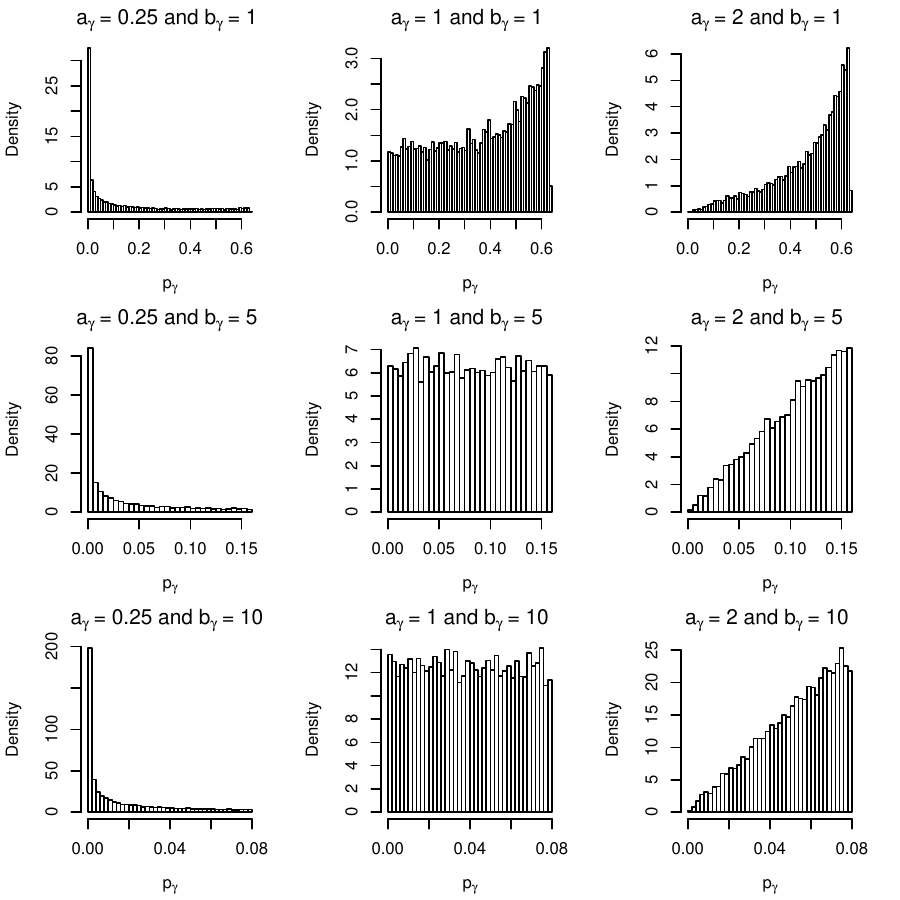} 
 \end{center}
 \vspace{-0.5cm}
 \caption[Distribution of the prior probability of non-shrinkage of the local basis coefficients, $p_\gamma := Pr(\gamma_{i,l} \neq 0 \mid K_\gamma)$, for different choices of Pareto hyperparameters $a_\gamma$ and $b_\gamma$.]{Distribution of the prior probability of non-shrinkage of the local basis coefficients, $p_\gamma := Pr(\gamma_{i,l} \neq 0 \mid K_\gamma)$, for different choices of Pareto hyperparameters $a_\gamma$ and $b_\gamma$.}
 \label{distributionpgamma}
\end{figure}

The hyperparameters in Equations (\ref{Kthetaprior})-(\ref{Kgammaprior}) have a key role in defining the prior (and posterior) probability of shrinkage of the periodic and local basis coefficients. Depending on the data and context, different degrees of sparsity might be desired, thus requiring careful tuning of these pa\-ra\-me\-ters. For simplicity, we focus here on the impact that different choices of $a_\gamma$ and $b_\gamma$ have on the prior probability of shrinkage of the local basis coefficients $\gamma_{i,l}$'s. \\
\indent With no loss of generality, assume $\bZ_l^\top \blambda_i = 0$ so that $\tilde{\gamma}_{i,l} \mid \bZ_l, \blambda_i \sim N(0, 1)$. Conditional on the latent thresholds, one can easily derive the probability that coefficient $\gamma_{i,l}$ is not switched off as 	
	\begin{equation}\label{conditional_prob_shrinking}
	P(\gamma_{i,l} \neq 0 \mid \varpi^*_{i,l}) = 2 \{1 - \Phi(\varpi^*_{i,l})\},
	\end{equation}
where $\Phi(\cdot)$ denotes the standard normal cumulative density function. By marginalizing over the prior distribution of the thresholds, we obtain 
	\begin{equation}\label{unconditional_prob_shrinking}
	p_\gamma := P(\gamma_{i,l} \neq 0 \mid K_\gamma) = \int_0^{K_\gamma} 2 \{1 - \Phi(\varpi^*_{i,l})\} \times \frac{1}{K_\gamma} d\varpi^*_{i,l},
	\end{equation}
which corresponds to the prior probability of non-shrinkage conditional on $K_\gamma$. Note that $p_\gamma$ is neither protein-dependent ($i$) nor basis-dependent ($l$). Integral (\ref{unconditional_prob_shrinking}) is not available in closed form, but can be evaluated via numerical integration. If we denote with $\gamma_{i,l}^*$ the indicator for the decision of not shrinking parameter $\gamma_{i,l}$, then the $\gamma^*_{i,l}$'s are independent and identically distributed ($i.i.d$) random variables with common probability of success $p_\gamma$,
	\begin{equation}\label{disitrib_indicators}
	\gamma^*_{i,l} := \mathbbm{1}(\gamma_{i,l} \neq 0)  \mid p_\gamma \mathop{\sim}^{i.i.d.} \text{Bernoulli}(p_\gamma).
	\end{equation}
Though separate decisions of shrinkage are to be taken on each parameter $\gamma_{i,l}$, the different cases are treated in unison within a framework of exchangeability. \\
\indent To evaluate the dependence of $p_\gamma$ to hyperparameters $a_\gamma$ and $b_\gamma$, we generated independent re\-a\-li\-za\-tions of $K_\gamma$ from the prior distribution (\ref{Kgammaprior}) given a particular choice of $a_\gamma$ and $b_\gamma$. For each of these realizations, we evaluated integral (\ref{unconditional_prob_shrinking}) and constructed the histogram of the so-obtained $p_\gamma$'s. Figure~\ref{distributionpgamma} shows the distribution of $p_\gamma$ for different choices of $a_\gamma$ and $b_\gamma$, with $b_\gamma$ varying along the rows and $a_\gamma$ along the columns. It emerges clearly that $b_\gamma$ defines an upper bound on the prior probability of non-shrinkage. Larger choices of $b_\gamma$ determine a smaller upper bound on $p_\gamma$ (the upper bound of the $x$-axis decreases by moving along the rows), thus effectively favoring a more sparse structure. For any given value of $b_\gamma$, small (large) values of $a_\gamma$ tend to favor smaller (larger) $p_\gamma$ whereas for $a_\gamma \approx 1$ the prior distribution of $p_\gamma$ becomes
	\begin{equation}\label{uniform_prio_pgamma}
	p_\gamma \mathop{\sim}^{approx} \text{Uniform}(0, U_{p_\gamma}),
	\end{equation}
where $U_{p_\gamma}$ is the upper bound on the distribution of $p_\gamma$. This result is true for larger choices of $b_\gamma$ in particular. 

A context where we expect high sparsity with, say, 90\% thresholding implies a fairly high choice of $b_\gamma$, and a value of $b_\gamma = 10$ or
above leads to a marginal sparsity probability exceeding 0.92. Unless the context involves substantive information to suggest favoring smaller or larger degrees of expected sparsity, an approximately uniform prior with $a_\gamma = 1$ and $b_\gamma = 5$ or $10$ is a good default for $p_\gamma$. A similar reasoning follows for $p_\theta$.

\subsection{Period detection}
The LTM on the periodic basis coefficients eases the identification of those proteins that are more likely to be periodically expressed.  Denote with $TS$ the total number of thinned posterior samples post-burn-in obtained by running a Markov-Chain-Monte-Carlo (MCMC) algorithm to update the model parameters, i.e. $TS = \frac{\text{Tot. }\# \text{ runs } - \text{burn-in}}{\text{thin}}$ (see Section \ref{Post} for details). We can easily derive the posterior probability of any simple periodicity (4, 6, 8 hours, etc.) by counting the proportion of posterior samples for which $\{\theta_{i, 2m - 1}, \theta_{i, 2m }\}$ are {\it{not}} shrunk to zero while the remaining $\btheta_s$'s are switched off. For example, the posterior probability that protein $i$ is  circadian can be computed as
	\begin{equation}\label{prob_circa}
	P(\text{Protein $i$ is  circadian}) = \frac{1}{TS} \sum_{g = 1}^{TS}\mathbbm{1}(\{\theta^{(g)}_{i,l}\}_{l = 1}^{2q - 2} \equiv \bzr   \text{ and } \{\theta^{(g)}_{i,2q - 1}, \theta^{(g)}_{i, 2q}\} \neq \bzr )
	\end{equation}
If we were interested in quantifying the probability of a protein being periodically expressed without making any specific reference to its period, we could simply count the proportion of posterior samples for which any pair $\{\theta_{i, 2m - 1}, \theta_{i, 2m}\}$ is {\it{not}} shrunk whereas the remaining parameters are switched off. In symbols,
      \begin{equation}\label{prob_periodic}
	P(\text{Protein $i$ is periodic}) = \frac{1}{TS} \sum_{g = 1}^{TS}\mathbbm{1}\left\{ 
	\begin{matrix}
	& [(\theta^{(g)}_{i, 1}, \theta^{(g)}_{i, 2}) \neq \bzr  \text{ and } (\theta^{(g)}_{i,l})_{l = 3}^{q} \equiv \bzr ]  \text{ or } \\
	& [(\theta^{(g)}_{i, 3}, \theta^{(g)}_{i, 4}) \neq \bzr  \text{ and } (\theta^{(g)}_{i,l})_{l \in\{1, 2, 4, \dots, q\}} \equiv \bzr ] \text{ or } \\
	& \vdots \\
	& [(\theta^{(g)}_{i,l})_{l = 1}^{2q - 2} \equiv \bzr \text{ and } (\theta^{(g)}_{i, 2q - 1}, \theta^{(g)}_{i, 2q}) \neq \bzr ]
	\end{matrix} \right\}
	\end{equation}
Biologists are interested in identifying clock proteins without incurring into too many false discoveries. Then, we need to compile a list of proteins for which the hypothesis of 24 hours periodicity is probably true, and we want the list to be as large as possible while bounding the rate of false discoveries by some threshold, say $k^*$. We can rank the proteins according to increasing values of $\beta_i = 1 - \text{Pr}(\text{Protein } i \text{ is  circadian})$ and declare all proteins with $\beta_i$ below a threshold, $\kappa$, as clock-controlled proteins
	\begin{equation}\label{clock_indicator}
	\beta_i^* = \mathbbm{1}(\beta_i \leq \kappa),
	\end{equation}
where $\beta_i^*$ is an indicator for the decision to report protein $i$ as circadian. \citep{Muller2004} show that (\ref{clock_indicator}) is the optimal decision rule under several loss functions that combine false negative and false discovery counts and/or rates, and the choice of the loss function determines the specific value of $\kappa$. In addition, the authors show that the result is true for any probability model with non-zero prior probability for periodic and non-periodic expression. In particular, the probability model can include dependence across proteins. \\
\indent Given the data, the expected number of false discoveries is 
	$$C(\kappa) = \sum_i \beta_i \mathbbm{1}[\beta_i \leq \kappa]$$
since $\beta_i$ is the conditional probability that identifying protein $i$ as  circadian creates a type I error. Hereafter we follow \cite{Newton2004} and choose a data-dependent $\kappa \leq 1$ as large as possible such that $C(\kappa) / \vert J \vert \leq k^*$, where $\vert J \vert > 0$ is the size of the list. So, $C(\kappa) / \vert J \vert$ is the expected rate of false discoveries given the data. 

\subsection{Inference on phase and amplitude}
In addition to period estimation, phase and amplitude of rhythmic transcripts must be accurately estimated. Grouping rhythmic transcripts by phase may suggest a common underlying regulatory mechanism. Also, the most robust cyclic proteins can be identified by amplitude. By making use of standard results from Fourier analysis, any simply periodic function $A \cos(\frac{2\pi}{w} t - \psi)$ can be expressed in an essentially unique manner as
	\[A \cos\left(\frac{2\pi}{w} t - \psi\right) = A_1 \sin\left(\frac{2\pi}{w} t\right) + A_2 \cos\left(\frac{2\pi}{w} t\right) \]
The function above is said to have amplitude $A$ and phase shift $\psi$. Therefore, the de-trended and centered true signal for protein $i$ at time $t_j$ can be written as
	\begin{align}\label{sinusoidal_decomp}
	f_i(t_j) &= \sum_{m = 1}^q A_{i,m} \cos\left( \frac{2\pi}{w_m} t_j - \psi_{i,m} \right) \\
		    &= \sum_{m = 1}^q \left[\theta_{i, 2m -1} \sin\left(\frac{2\pi}{w_m} t_j\right) + \theta_{i, 2m} \cos\left(\frac{2\pi}{w_m} t_j\right)\right], \label{sinu_used}
	\end{align}
where $A_{i,m}$ and $\psi_{i,m}$ denote the amplitude and phase of the oscillation with period length $w_m$. We need to identify $A_{i,m}$ and $\psi_{i,m}$. If we write out 
	\[\cos\left(\frac{2\pi}{w_m} t_j - \psi_{i,m}\right) = \sin\left(\frac{2\pi}{w_m} t_j\right)\sin\psi_{i,m} + \cos\left(\frac{2\pi}{w_m} t_j\right)\cos\psi_{i,m}\]
we see that we must have
	\[A_{i,m} \sin \psi_{i,m} = \theta_{i, 2m -1} \qquad \text{and} \qquad A_{i,m} \cos \psi_{i,m} = \theta_{i, 2m}\]
Therefore, we get
	\begin{align}\label{formula_for_amplitude}
	& A_{i,m}  = \sqrt{\theta_{i, 2m-1}^2 + \theta_{i, 2m}^2}, \quad \text{and} \\ 
	& \psi_{i,m}  = \tan^{-1} \left(\frac{\theta_{i, 2m - 1}}{\theta_{i, 2m}}\right). \label{formula_for_phase}
	\end{align}
The sets of period, amplitude, and phase $\{w_m, A_{i,m}, \psi_{i,m}\}$,  $m = 1, \dots, q$, provide a complete description of the true process $f_i(t_j)$ underlying the observed oscillation. 

\section{Posterior update}
\label{Post}
Given the observed data $\bY = \{\by_i\}_{i = 1}^p$, we wish to infer the periodic basis functions coefficients $\{\btheta_i\}_{i = 1}^p$, the local basis functions coefficients $\{\bgamma_i\}_{i = 1}^p$, the factor loading matrix $\bLambda$, the $T \times k$ matrix of latent factors $\beeta$, and all hyperparameters. We use Gibbs sampling by successively drawing samples from the full conditional distributions of each parameter in turn, given all other parameters.\\

The conditional distribution of $\bY$ implied by (\ref{yi_model}) is
	\begin{equation}\label{model_Y}
	\bY \vert \bB, \bC, \bTheta, \bGamma, \beeta, \bLambda, \bSigma = \prod_{i = 1}^p \text{N}(\by_i \vert \bB \btheta_i + \bC \bgamma_i + \beeta\blambda_i, \sigma_i^2 \bI_T),
	\end{equation}
and the likelihood function is
	   \begin{eqnarray}
         \label{likelihood_fct}
       &&P(\bY , \bTheta, \bGamma, \tilde{\bTheta}, \tilde{\bGamma}, \bSigma, \beeta, \bLambda, \bphi, \btau, \bvarpi, \bvarpi^*)= \\
        \nonumber &&\prod_{i = 1}^p  \Big\{\text{N}(\by_i \vert \bB\btheta_i  + \bC\bgamma_i + \beeta
\blambda_i, \sigma_i^2 \bI_T) \text{Ga}(\sigma_i^{-2}\vert a_\sigma, b_\sigma)\times \\
   \nonumber  &&N_k\left[\blambda_i^\top \vert \bZero, \bD_i(\bphi,
\btau)\right] p(\bphi\vert\rho) p(\btau|a_1, a_2) \prod_{j = 1}^T N_k (\beeta_{j} \vert
\bzr, \bI_k)\times \\
   \nonumber  &&  N_{2q}(\tilde{\btheta}_i \vert \bW\blambda_i, \mathbbm{V}\text{ar}(\balpha_i^\theta))\times N_{\tilde{T}}(\tilde{\bgamma}_i \vert \bZ\blambda_i, \mathbbm{V}\text{ar}(\balpha_i^\gamma)) \times p(K_\theta) \times p(K_\gamma) \times\\
\nonumber && \prod_{j = 1}^{2q}N_{k}(\bW_j \vert \bzr, \bI_k) \times \prod_{j = 1}^{\tilde{T}}N_{k}(\bZ_j \vert \bzr, \bI_k) \times p(\bvarpi) \times p(\bvarpi^*)\Big\},
   \end{eqnarray}
   where $p(\bphi\vert\rho)$ and $p(\btau|a_1, a_2)$ are the densities of prior distributions induced by MGPSP on vectors of all $\left\{\phi_{ih}\right\}_{i=1, \ldots, p;~h=1, \ldots,k}$ and all $\left\{\tau_h\right\}_{h=1, \ldots,k}$, respectively, and $p(\bvarpi)$ and $p(\bvarpi^*)$ are the densities of prior distributions induced on vectors of all $\{\varpi_{i,m}\} _{i = 1, \ldots, p;~m = 1, \ldots, q}$ and $\{\varpi_{i,l}^*\} _{i = 1, \ldots, p;~l = 1, \ldots, \tilde{T}}$, respectively. 
   
 In what follows we  use ``--'' to denote the ``rest'' of the model, i.e. all random variables not explicitly mentioned in the current state of the Markov Chain. Using the introduced notations we describe a MCMC algorithm for simulation of the full joint posterior distribution of the model parameters. \\
 
  \begin{itemize}
  \item{\it{Update of $\bW$}:} We place a conjugate normal prior on the columns of the $k \times 2q$ matrix $\bW^\top$, so $\bW_{l}  \sim N_k(\bzr, \bI), l = 1, \dots, 2q$. This is equivalent to a prior on the rows of matrix $\bW$, $ \bW_l^\top$. Conditioning on the current estimate of $\tilde{\theta}_{i,l} \sim N(\blambda_i^\top \bW_{l}, 1)$ and other model parameters, the posterior update of $\bW_{l}$ is 
  	\begin{equation*}\label{posterior_W}
	\bW_{l} \mid - \sim N_k\left(\left(\sum_{i = 1}^p \blambda_i \blambda_i^\top + \bI \right)^{-1}\left(\sum_{i = 1}^p \tilde{\theta}_{i,l}\blambda_i\right), \left(\sum_{i = 1}^p \blambda_i \blambda_i^\top + \bI \right)^{-1}\right)
 	\end{equation*}
	
	  \item{\it{Update of $\bZ$}:} We place a conjugate normal prior on the columns of the $k \times \tilde{T}$ matrix $\bZ^\top$, so $\bZ_{l} \sim N_k(\bzr, \bI), l = 1, \dots, \tilde{T}$. This is equivalent to a prior on the rows of matrix $\bZ$, $ \bZ_l^\top$. Conditioning on the current estimate of $\tilde{\gamma}_{i,l} \sim N(\blambda_i^\top \bZ_{l}, 1)$ and other model parameters, the posterior update of $\bZ_{l}$ is 
  	\begin{equation*}\label{posterior_Z}
	\bZ_{l} \mid - \sim N_k\left(\left(\sum_{i = 1}^p \blambda_i \blambda_i^\top + \bI \right)^{-1}\left(\sum_{i = 1}^p \tilde{\gamma}_{i,l}\blambda_i\right), \left(\sum_{i = 1}^p \blambda_i \blambda_i^\top + \bI \right)^{-1}\right)
 	\end{equation*}
	\item{\it{Update of $\blambda_i^\top$}:} We place a MGPSP on row $i$ of \bLambda (equivalently, column $i$ of $\bLambda^\top$) as in (\ref{mgpsp}). The likelihood contribution factorizes as
	\begin{align}\label{likelihood_lambda}
	L(\blambda_i \vert \tilde{\bTheta}, \bTheta, \tilde{\bGamma}, \bGamma, \beeta, \bSigma, \bW, \bZ) &\propto N_T(\by_i \vert \bB\btheta_i + \bC\bgamma_i + \beeta \blambda_i, \sigma_i^2 \bI) \times \\ \nonumber
	& \times N(\tilde{\btheta}_i \vert \bW \blambda_i , \mathbbm{V}\text{ar}(\balpha_i^\theta))  \times N(\tilde{\bgamma}_i \vert \bZ \blambda_i , \mathbbm{V}\text{ar}(\balpha_i^\gamma))
	\end{align}
We assume $\mathbbm{V}\text{ar}(\balpha_i^\theta) = \bI_{2q}$ and $\mathbbm{V}\text{ar}(\balpha_i^\gamma) = \bI_{\tilde{T}}$. The posterior update of $\blambda_i$ is
	\begin{eqnarray}
	 \label{posterior_lambda_i}
  \blambda_i\mid \al &\sim&
\text{N}_{k}\left(\bV_{\blambda_i}\bM_{\blambda_i},
\bV_{\blambda_i}\right), \quad i = 1, \dots, p, \quad \text{ where }\\
\nonumber&&~\bM_{\blambda_i}= \sigma_i^{-2}\beeta^\top(\by_i - \bB\btheta_i - \bC\bgamma_i) + \bW^\top \tilde{\btheta}_i + \bZ^\top \tilde{\bgamma}_i\\
\nonumber&&~ \bV_{\blambda_i}=\left(\tfrac{1}{\sigma_i^2}\beeta^\top\beeta + \bW^\top \bW + \bZ^\top \bZ + \bD_i^{-1}\right)^{-1}
\end{eqnarray}

\item {\it{Update of $\tilde{\btheta}_i$}:} We sample the conditional posterior $p(\tilde{\btheta}_i \mid \al)$ sequentially for $i = 1, \dots, p$ using a Metropolis-Hastings (MH) sampler conditional on the other model parameters. The MH proposal originates from a non-thresholded version of the model. Fixing $\mathbbm{1}(\vert \vert \tilde{\btheta}_{i,m} \vert \vert \geq \varpi_{i,m}) \equiv 1$ for $m = 1, \dots, q$, we take the proposal distribution to be $N(\tilde{\btheta}_i \mid \boldsymbol{m}_i, \boldsymbol{M}_i)$ with
	\begin{align*}
	\boldsymbol{M}_i &= \left(\sigma_i^{-2} \bB^\top \bB + \bI_{2q}\right)^{-1},  \\
	\boldsymbol{m}_i &= \boldsymbol{M}_i \times (\sigma_i^{-2} \bB^\top \tilde{\by}_i + \bW \blambda_i)
	\end{align*}
with $ \tilde{\by}_i = \by_i - \bC\bgamma_i - \bbeta \blambda_i$. The candidate is accepted with probability
	\begin{equation*}\label{MH_acc_theta}
	\alpha(\tilde{\btheta}_i, \tilde{\btheta}^*_i) = \min\left\{1, \frac{N(\by_i \mid \bB\btheta_i^* + \bC \bgamma_i + \bbeta \blambda_i, \sigma_i^2\bI_T)N(\tilde{\btheta}_i^* \mid \bW\blambda_i, \bI) N(\tilde{\btheta}_i \mid \boldsymbol{m}_i, \boldsymbol{M}_i)}{N(\by_i \mid \bB\btheta_i + \bC \bgamma_i + \bbeta \blambda_i, \sigma_i^2\bI_T)N(\tilde{\btheta}_i \mid \bW\blambda_i, \bI) N(\tilde{\btheta}_i^* \mid \boldsymbol{m}_i, \boldsymbol{M}_i)} \right\}
	\end{equation*}
where $\tilde{\btheta}_i$ ($\btheta_i$) is the current estimate and $\tilde{\btheta}_{i,m}^*$ (${\btheta}^*_{i,m} = \tilde{\btheta}_{i,m}^*\mathbbm{1}(\vert \vert \tilde{\btheta}_{i,m}^* \vert \vert \geq \varpi_{i,m})$) is the candidate, with $\btheta_i^* = \{{\btheta}^*_{i,m}\}_{m = 1}^q$. 

\item {\it{Update of $\tilde{\bgamma}_i$}:} We sample the conditional posterior $p(\tilde{\bgamma}_i \mid \al)$ sequentially for $i = 1, \dots, p$ via MH with proposals obtained from a non-thresholded version of the model. Fixing $\mathbbm{1}(\vert \tilde{\gamma}_{i,l} \vert \geq \varpi^*_{i,l}) \equiv 1$ for $l = 1, \dots, \tilde{T}$, we take the proposal distribution to be $N(\tilde{\bgamma}_i \mid \boldsymbol{n}_i, \boldsymbol{N}_i)$ where
	\begin{align*}
	\boldsymbol{N}_i &= \left(\sigma_i^{-2} \bC^\top \bC + \bI_{\tilde{T}}\right)^{-1},  \\
	\boldsymbol{n}_i &= \boldsymbol{N}_i \times (\sigma_i^{-2} \bC^\top \tilde{\by}_i + \bZ \blambda_i)
	\end{align*}
with $ \tilde{\by}_i = \by_i - \bB\btheta_i - \bbeta \blambda_i$. The MH acceptance probability is
	\begin{equation*}\label{MH_acc_gamma}
	\alpha(\tilde{\bgamma}_i, \tilde{\bgamma}^*_i) = \min\left\{1, \frac{N(\by_i \mid \bB\btheta_i + \bC \bgamma_i^* + \bbeta \blambda_i, \sigma_i^2\bI_T)N(\tilde{\bgamma}_i^* \mid \bZ\blambda_i, \bI) N(\tilde{\bgamma}_i \mid \boldsymbol{n}_i, \boldsymbol{N}_i)}{N(\by_i \mid \bB\btheta_i + \bC \bgamma_i + \bbeta \blambda_i, \sigma_i^2\bI_T)N(\tilde{\bgamma}_i \mid \bZ\blambda_i, \bI) N(\tilde{\bgamma}_i^* \mid \boldsymbol{n}_i, \boldsymbol{N}_i)} \right\}
	\end{equation*}
where $\tilde{\bgamma}_i$ ($\bgamma_i$) is the current estimate and $\tilde{\bgamma}_{i,l}^*$ (${\bgamma}^*_{i,l} = \tilde{\bgamma}_{i,l}^*\mathbbm{1}(\vert \tilde{\bgamma}_{i,l}^* \vert \geq \varpi_{i,l}^*)$) is the candidate, with $\bgamma_i^* = \{{\bgamma}^*_{i,l}\}_{l = 1}^{\tilde{T}}$. 

\item  {\it{Update of $\varpi_{i,m}$}:} The update can be performed via Gibbs sampling conditioning on the current estimate of $\tilde{\btheta}_{i,m} = \{\tilde{\theta}_{i,2m-1}, \tilde{\theta}_{i,2m}\}^\top$ and the other model parameters for $i = 1, \dots, p$ and $m = 1, \dots, q$. If $\vert\vert \tilde{\btheta}_{i,m} \vert \vert > K_\theta$ (the upper bound of the uniform prior on $\varpi_{i,m}$), the posterior update of $\varpi_{i,m}$ is 
	\begin{equation*}\label{post_latent_firstcase_theta}
	\varpi_{i,m} \mid \al \sim \text{Unif}(0,  K_\theta). 
	\end{equation*}
Otherwise, sample
	\[ \varpi_{i,m} \mid \al \sim \left\{ 
          \begin{array}{l l}
                \text{Unif}(0, \vert \vert \tilde{\btheta}_{i,m} \vert \vert) & \quad \text{with probability} \quad \pi^*\\
                \text{Unif}(\vert \vert \tilde{\btheta}_{i,m} \vert \vert, K_\theta) & \quad \text{with probability} \quad 1 - \pi^*,
           \end{array} \right.\] 
   with
   	\begin{align*}\label{pi_star}
	\pi^* &= \frac{ A } { A + D}, \\
	A &= N(\by_i \mid \bB_{-m}\btheta_{i, -m} + \bB_m\tilde{\btheta}_{i,m} + \bC\bgamma_i + \beeta\blambda_i, \sigma^2_i\bI_T) \times \vert \vert \tilde{\btheta}_{i,m} \vert \vert, \\
	D &=  N(\by_i \mid \bB_{-m}\btheta_{i, -m} + \bC\bgamma_i + \beeta\blambda_i, \sigma^2_i\bI_T) \times (K_\theta - \vert \vert \tilde{\btheta}_{i,m} \vert \vert ),
	\end{align*} 
with $N(\by_i \mid \boldsymbol{m}, \boldsymbol{v})$ denotes the Gaussian density function with mean $\boldsymbol{m}$ and covariance matrix $\boldsymbol{v}$ evaluated at $\by_i$.
Matrix $\bB_{-m}$ ($\btheta_{i,-m}$) corresponds to the matrix of periodic bases (vector of periodic basis coefficients) with columns (components) $m = \{2m-1, 2m\}$ excluded. Instead, $\bB_m$ ($\tilde{\btheta}_{i,m}$) denotes the $\{2m-1, 2m\}$-th columns of matrix $\bB$ (the $\{2m-1, 2m\}$-th components of $\tilde{\btheta}_i$).


\item  {\it{Update of $\varpi^*_{i,l}$}:} The update can be performed via Gibbs sampling conditioning on the current estimate of $\tilde{\gamma}_{i,l}$ and the other model parameters for $i = 1, \dots, p$ and $l = 1, \dots, \tilde{T}$. If $\vert \tilde{\gamma}_{i,l} \vert > K_\gamma$ (the upper bound of the uniform prior on $\varpi^*_{i,l}$), the posterior update of $\varpi^*_{i,l}$ is  
	\begin{equation*}\label{post_latent_firstcase}
	\varpi^*_{i,l} \mid \al \sim \text{Unif}(0, K_\gamma). 
	\end{equation*}
Otherwise, sample
	\[ \varpi^*_{i,l} \mid \al \sim \left\{ 
          \begin{array}{l l}
                \text{Unif}(0, \vert \tilde{\gamma}_{i,l} \vert) & \quad \text{with probability} \quad \pi^*\\
                \text{Unif}(\vert \tilde{\gamma}_{i,l} \vert, K_\gamma) & \quad \text{with probability} \quad 1 - \pi^*,	
           \end{array} \right.\]
   with
   	\begin{align*}\label{pi_star}
	\pi^* &=  \frac{ E } { E + F}, \\
	E &= N(\by_i \mid \bB\btheta_i + \bC_{-l}\bgamma_{i,-l} + \bC_l\tilde{\gamma}_{i,l} + \beeta\blambda_i, \sigma^2_i\bI_T) \times \vert \tilde{\gamma}_{i,l} \vert, \\
	F &= N(\by_i \mid \bB\btheta_i + \bC_{-l}\bgamma_{i,-l} + \beeta\blambda_i, \sigma^2_i\bI_T) \times (K_\gamma - \vert \tilde{\gamma}_{i,l} \vert) 
	\end{align*}
Matrix $\bC_{-l}$ ($\bgamma_{i,-l}$) corresponds to the matrix of local bases (vector of local basis coefficients) with column (component) $l$ excluded. Instead, $\bC_l$ ($\tilde{\gamma}_{i,l}$) denotes the $l$-th column of matrix $\bC$ (the $l$-th component of $\tilde{\bgamma}_i$).
  \end{itemize}
  
 Further,
  \begin{eqnarray*}
 \label{post_ktheta}
 K_\theta \mid \al &\sim& \text{Pareto}\left(a_\theta + pq, \max\{b_\theta, \max_{i, m} \{\varpi_{i,m}\}_{i = 1, m = 1}^{p, q}\}\right); \\
  \label{post_kgamma}
 K_\gamma \mid \al &\sim& \text{Pareto}\left(a_\gamma + p\tilde{T}, \max\{b_\gamma, \max_{i, l} \{\varpi^*_{i,l}\}_{i = 1, l = 1}^{p, \tilde{T}}\}\right); \\
\label{post_sigma}
 \sigma_i^{-2} \mid \al &\sim& \text{Ga}\left(a_\sigma + \frac{T}{2}, b_\sigma +
\frac{\vert\vert \by_i - \bB\btheta_i - \bC \bgamma_i - \beeta\blambda_i \vert\vert^2}{2}\right), \quad i = 1, \dots, p;\\
\label{post_mod_eta}
\beeta_{j}\mid \al &\sim& N_k\left[\bV_{\beeta_{j}}\bM_{\beeta_{j}}, \bV_{\beeta_{j}}\right], \quad j = 1, \dots, T, \quad \text{ where }\\
\nonumber&&~\bM_{\beeta_{j}}=\bLambda^\top \bSigma^{-1}(\by^{(j)}- \bTheta\boldsymbol{b}_j - \bGamma \boldsymbol{c}_j ),\\
\nonumber &&~\bV_{\beeta_{j}}=(\bI_k+\bLambda^\top\bSigma^{-1}\bLambda)^{-1}; \\
\label{post_phi} 
\phi_{ih} \mid \al & \sim & \text{Ga}\left(\frac{\rho + 1}{2},  \frac{\rho + \tau_h \lambda^2_{ih}}{2}\right),  \quad i = 1, \dots, p \quad \text{and} \quad h = 1, \dots, k;\\
\label{post_zeta} 
\zeta_1 \mid \al &\sim& \text{Ga}\left(a_1 + \frac{pk}{2}, 1 + \frac{1}{2} \sum_{l=h}^{k}\tau_l^{(1)}\sum_{i=1}^p \phi_{il}\lambda_{il}^2\right); \\
\label{post_zeta_more} \zeta_h \mid\al &\sim& \text{Ga}\left(a_2 + \frac{p}{2}(k-h+1), 1 + \frac{1}{2} \sum_{l=1}^{k}\tau_l^{(h)}\sum_{i=1}^p \phi_{il}\lambda_{il}^2\right), \\
\nonumber && \text{for} \quad h\geq 2, \quad \text{where} \quad \tau_l^{(h)} = \prod_{t=1, t \neq h}^{l}\zeta_t \quad \text{for} \quad h=1, \dots, k.
\end{eqnarray*}

\section{Simulation studies}
\label{lf_simulation}
\subsection{Dependence across measurements}
\label{Case1:dependence}
\begin{figure}[h!]
 \begin{center}
 \includegraphics[width=16cm, height = 18cm]{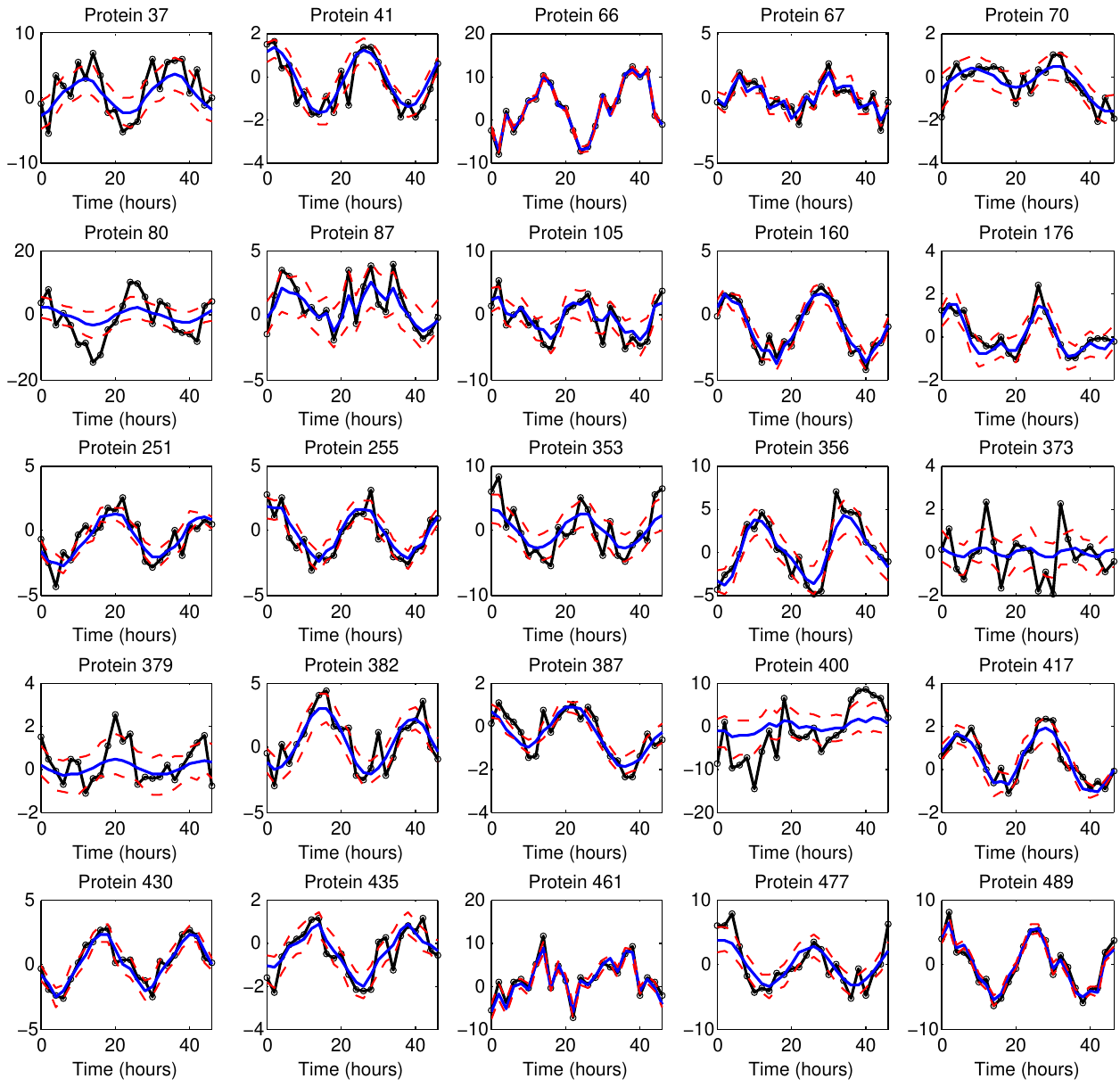} 
 \end{center}
 \vspace{-2.5cm}
 \caption[True (black) and inferred (blue) trajectories of the 25 truly circadian variables in the simulation study of dependence across variables]{True (black) and inferred (blue) trajectories of the 25 truly circadian variables in the simulation study of Section \ref{Case1:dependence}, with the horizontal axis corresponding to time. Red lines are the pointwise 95\% credible intervals. }
 \label{Case1dependence}
\end{figure}

\begin{figure}[h!]
 \begin{center}
 \includegraphics[width=16cm, height = 8cm]{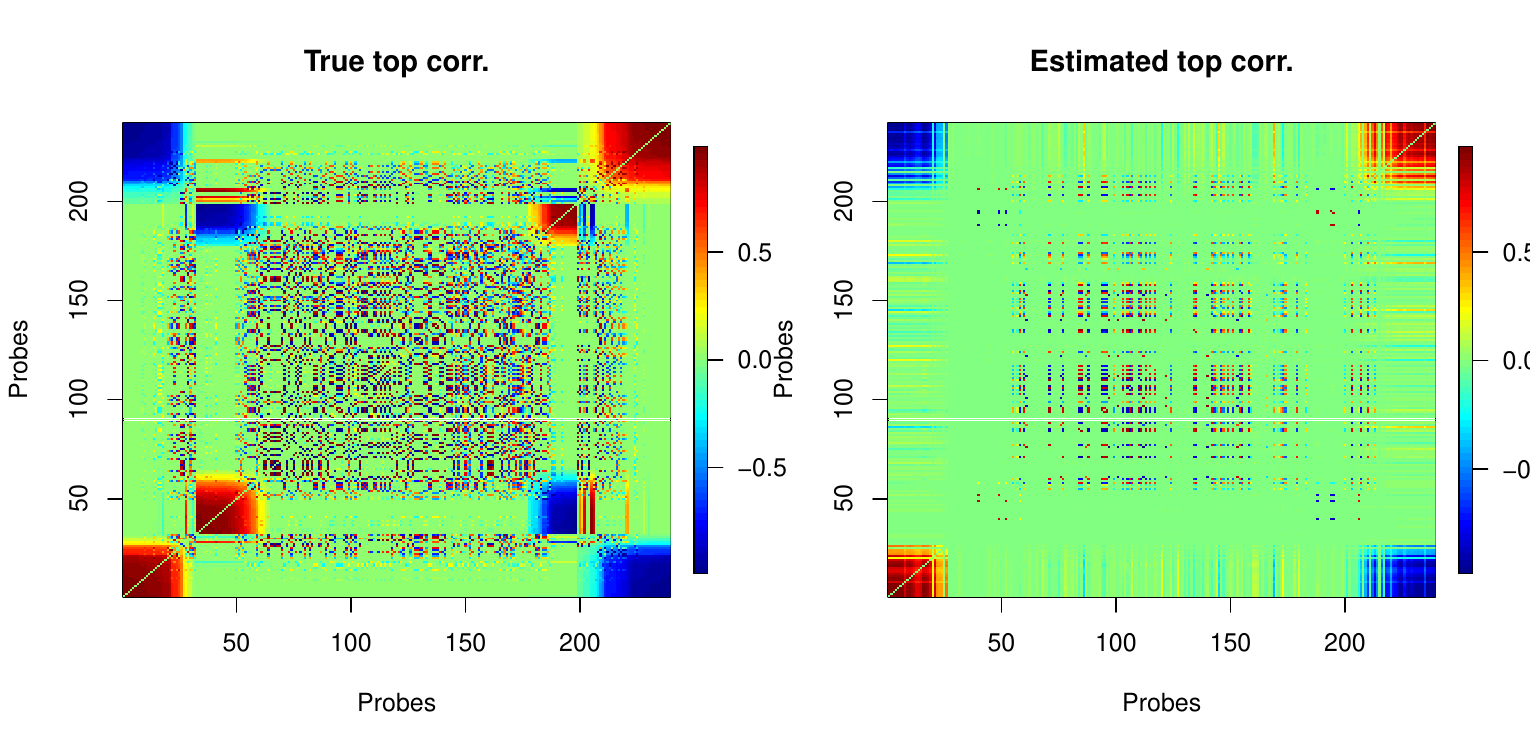} 
 \end{center}
 \vspace{-0.5cm}
 \caption[Comparison between true and estimated correlation structure among probes with true pair-wise correlation of or above (below) $0.90$ (-0.90)]{Comparison between true and estimated correlation structure among probes with true pair-wise correlation of or above (below) $0.90$ (-0.90). To improve visibility, the diagonal of the correlation matrix is set equal to 0.}
 \label{sim_corr_struct}
\end{figure}

\begin{figure}[h!]
 \begin{center}
 \includegraphics[width=16cm, height = 8cm]{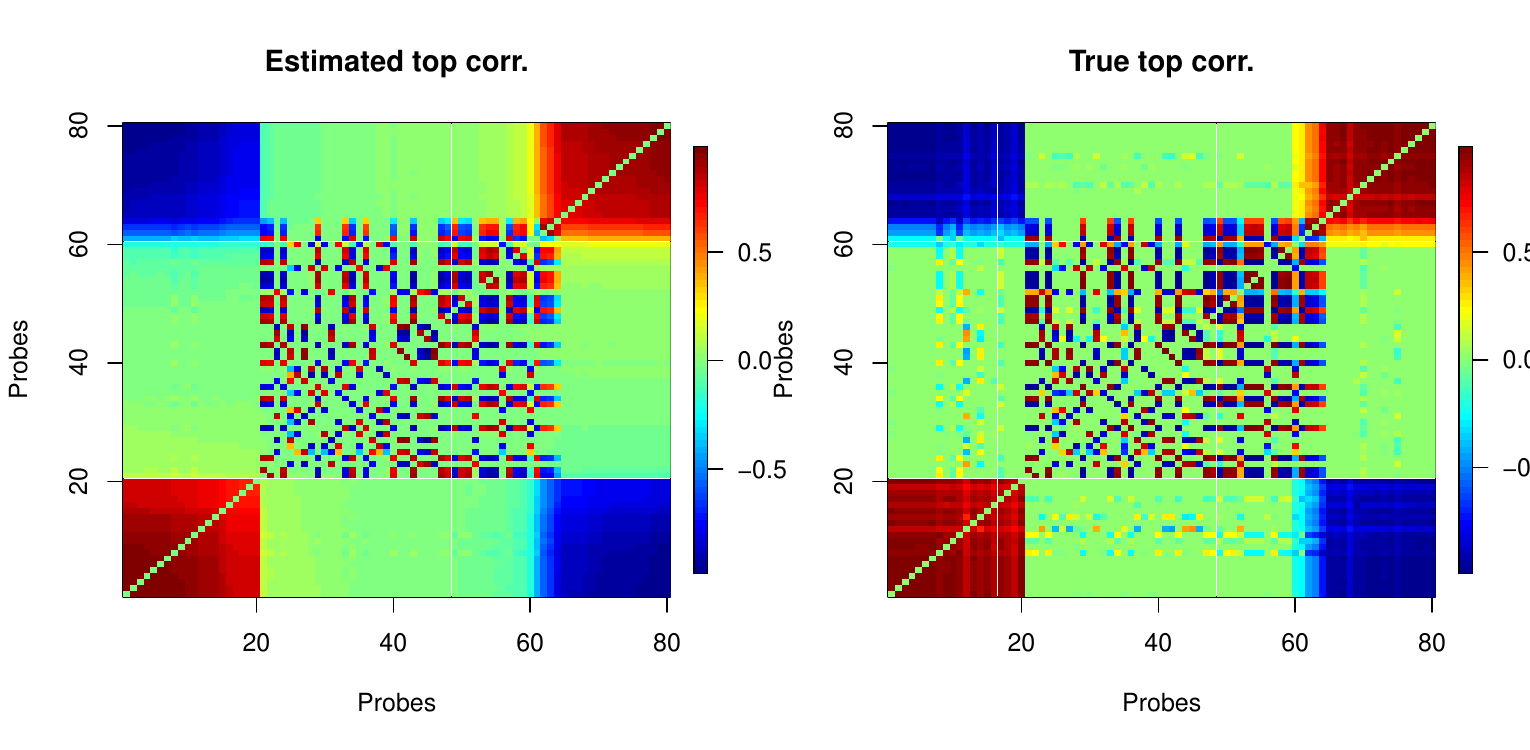} 
 \end{center}
 \vspace{-0.5cm}
 \caption[Comparison between estimated and true correlation structure among probes with estimated pair-wise correlation of or above (below) $0.80$ (-0.80)]{Comparison between estimated and true correlation structure among probes with estimated pair-wise correlation of or above (below) $0.80$ (-0.80). To improve visibility, the diagonal of the correlation matrix is set equal to 0.}
 \label{top_corr_estimated}
\end{figure}

We synthesized data from the model, and then used the above framework to infer the model parameters. We simulated $\by_i$, $i = 1, \dots, p = 500$, from a $T = 24$-dimensional normal distribution with mean $\bB\btheta_i + \bC\bgamma_i + \beeta\blambda_i$ and covariance matrix $\sigma_i^2 \times \bI_T$, with $\sigma_i^2 = 0.5 $ $\forall i$. The design matrix $\bB$ included the Fourier bases as specified in Section \ref{Motivation} with possible periods $\{4,6,8,12,24\}$ hours, thus $q = 5$, whereas $\tilde{T} = 10$ Gaussian kernels with common bandwidth $\psi = 25$ were chosen for the matrix of local bases $\bC$. The true number of factors was set equal to $k = 6$, and the number of non-zero elements in each column of $\bLambda$ were chosen linearly between $2 \times (10\log p)$ and $10 \log p + 1$. In practice, this resulted in a number of non-zero elements between 99 and 124 across the different columns of $\bLambda$. We randomly allocated the location of the zeros in each column and simulated the non-zero elements independently from a normal distribution with mean 0 and variance 9. The latent factors $\beeta$ were independently generated by sampling from a standard normal distribution. The $p \times q$ true latent thresholds for $\bTheta$ were independently generated from a Unif$(0,6)$ whereas the $p \times \tilde{T}$ latent thresholds for $\bGamma$ were independently generated from a Unif$(0, 10)$ to induce sparsity on $\bGamma$ and jitter the curves with only a few, time-localized deviations. The rows of $\bW$ ($\bZ$) were independently generated by sampling from a standard normal distribution, and the true values of the latent coefficients $\{\tilde{\btheta}_i, \tilde{\bgamma}_i\}_{i = 1}^p$ were generated by sampling from their prior distribution given the true values of $\bW, \bZ$, and $\bLambda$. \\
\indent We run the Gibbs sampler described in Section \ref{Post} for 50000 iterations with a burn-in of 20000, and collected every 5th sample to thin the chain. The hyperparameters $a_\sigma$ and $b_\sigma$ for $\sigma_i^{-2}$ were 1 and 0.5, respectively, while $\rho = 3$, $a_1 = 2.1$, $a_2 = 3.1$, $a_\theta = a_\gamma = 1$, $\beta_\theta = 5, \beta_\gamma = 10$ and used $k = 5$ as the starting number of factors. \\
\indent Of the 500 curves, 22.4\% exhibit simple periodicity with periods either $\{4,6,8,12,24\}$ hours and the remaining profiles either load on more than one Fourier basis or are pure noise. Only 25 of the 500 simulated profiles truly exhibit circadian expression. Therefore, the signal-to-noise ratio is quite weak in this dataset. Figure~\ref{Case1dependence} shows the estimated trajectories for the 25  circadian variables: the black line represents the true trajectory, the blue line represents the posterior mean estimate of $\bB\btheta_i + \bC \bgamma_i + \bbeta \lambda_i$, and the red dashed lines are the 95\% pointwise credible intervals of the same quantity. 
\indent Figure~\ref{sim_corr_struct} shows a comparison between the true correlation (left panel) and the estimated correlation structure (right panel). To improve visibility, the plot only reports probes that give rise to true pair-wise correlations of or above (below) $0.90$ (-0.90). The correlation structure seems overall slightly under-estimated: this is likely the effect of the MGPS prior on $\bLambda$, which tends to favor small (in magnitude) loadings, as opposed to the wide-support distribution ($N(0,9)$) used to generate the true non-zero elements on $\bLambda$. By examining the correlation matrix generated by probes with {\textit{estimated}} pair-wise correlation of or above (below) $0.80$ (-0.80), we notice an almost perfect match with their true correlation structure (Figure~\ref{top_corr_estimated}).
\indent When computing model-parameter summaries, one must address the fact that the state (zero or non-zero) of the periodic and local basis coefficients $\{\btheta_i, \bgamma_i\}$ may change between collection of samples (it should change between collection of samples if there is good mixing). When aggregating collection samples, reporting the posterior mean for these parameters could be misleading in that the non-zero estimates will bias the overall posterior mean, which is therefore unlikely to be exactly zero even when the true value of the parameter is zero. Table \ref{Case1dependence:table} reports the proportion of posterior samples for which $\btheta_m \equiv \{\theta_{i,2m - 1}, \theta_{i,2m}\}, m = 1, \dots, 4$ are estimated being equal to zero while $\btheta_5 \equiv \{\theta_{i,9}, \theta_{i, 10}\}$ are estimated being different from zero for the 25 circadian variables. We recall that $\btheta_5$ denotes the vector of coefficients of the sine/cosine bases with 24 hours period. The table also shows the estimated circadian probability (\ref{prob_circa}) and quantiles of the inferred phase and amplitude of the oscillation with period length of 24 hours, $\psi_{i,5}$ and $A_{i,5}$. 
\begin{table}[ht]
\caption[Simulation study as of Section \ref{Case1:dependence}: proportion of posterior samples for which parameter $\theta_{i,m}$ is estimated being equal to zero for $m \in \{1, 3, 5, 7\}$ whereas parameter $\theta_{i, 2q - 1}$ is not shrunk. Also reported is the estimated circadian probability, phase and amplitude of the oscillation with period length of 24 hours]{Simulation study as of Section \ref{Case1:dependence}: columns refer to the different periodic basis parameters and rows to the truly circadian variables. Columns 2-5 report the proportion of posterior samples for which $\btheta_m = \{\theta_{i,2m-1}, \theta_{i, 2m}\}$ are estimated being equal to zero for $m = 1, \dots, 4$, whereas column 6 reports the proportion of posterior samples for which $\btheta_5 = \{\theta_{i, 9}, \theta_{i, 10}\}$ are estimated being different from zero. Vector $\btheta_5$ contains the coefficients of the oscillation with period length of 24 hours. Column 7 reports the estimated probability of a protein being  circadian, i.e. the proportion of posterior samples with $\{\theta_{i,m}\}_{m = 1}^{2q - 2} \equiv 0$ and $\{\theta_{i,2q - 1}, \theta_{i, 2q}\} \neq \bzr$. The last two columns report the true amplitude and phase versus the [2.5, 50, 97.5]\% quantiles of estimated amplitude and phase for the oscillation with period length of 24 hours (brackets). Ranking is by posterior circadian probability.} 
\label{Case1dependence:table}
\begin{center}
\begin{footnotesize}
\begin{tabular}{c c c c  c c  c  c c} 
\hline\hline 
Protein & $\btheta_{i,1}$ &  $\btheta_{i,2}$ & $\btheta_{i,3}$ &  $\btheta_{i,4}$ & $\btheta_{i,5}$ & P(Circ) & $A_{i,5}$ (true/est) & $\psi_{i,5}$ (true/est)\\ [0.5ex] 
\hline 
66 & 100 & 100  &  100 & 100 & 100 &  1 & 4.84 [4.1, 4.5, 4.9] &  0.48 [0.47, 0.50, 0.52] \\[0.5ex] 
251 & 100 & 99.7 & 95.1 & 99.9  & 100  & 0.95 &  1.46 [1.4, 1.4, 1.4] & 1.53 [-1.28,  -1.28, -1.28]    \\[0.5ex] 
489 & 98.5 & 98.7 &  99.3 & 97.5 & 100 & 0.94 & 4.09 [3.22, 3.73, 4.46] & 0.35 [0.25, 0.38, 0.45] \\[0.5ex] 
387 & 96  & 97.3 & 100 & 100 & 99.9  & 0.93 & 0.75 [0.86, 0.86, 0.98] & 0.14 [-0.67, -0.67, -0.56]    \\[0.5ex] 
41 & 91.4  & 99 &  99.6 &  97.6 & 100 & 0.88  & 1.26 [1.09, 1.37, 1.66] &  0.54 [0.34, 0.49, 0.76] \\[0.5ex] 
417 & 87.4  & 99 & 100 & 100 & 100 & 0.87 & 	1.44 [1.23, 1.23, 1.23] & 1.23 [1.1, 1.1, 1.1]	 \\[0.5ex] 
382 & 95.8  &  93.6 & 94.1 & 94.4 & 100  & 0.80 &	1.38 [1.68, 2.29, 2.78] & 0.87 [0.58, 0.86, 1.1]				 \\[0.5ex] 
477 & 92.8  & 94.1 & 94.2 &  87.7 & 100 & 0.72 & 4.09 [2.19, 3.14, 4.26] &0.33 [0.01, 0.41, 0.65]\\[0.5ex] 
255 & 75.8 & 92.6 & 98.8 & 98.9 & 100 & 0.69 &  2.43 [1.73, 1.89, 2.4] & 0.26 [0.18, 0.44, 0.6]					\\[0.5ex]   
160 & 66.5 &  95.7  & 99.1 & 93.4  & 100 &0.59 & 1.85 [1.43, 1.7, 2.13] & 1.07 [0.65, 0.9, 1.12]       \\[0.5ex] 
70 & 94.9  & 98.1 &  99 &  98.3 & 81.4 & 0.57 & 0.92 [0.45, 0.7, 1.12] & -0.39 [-1.32, -0.83, 1.51]  \\[0.5ex] 
435 & 97.5  & 99.4 & 62.4 & 96.3 & 99.8 & 0.57 & 	1.48  [0.71, 0.73, 1.35] & 0.47 [-0.01, 0.35, 0.69]		\\[0.5ex] 
87 & 92.2 & 75.6  &  83.4 &  97.5 & 93.9 & 0.53 & 1.66 [0.77, 1.28, 1.94] & 1.3 [-1.56, -1.04, 1.54] \\[0.5ex] 
37 & 87.9 & 85 & 79 & 88.9 & 98.8  & 0.53  &  4.96 [1.44, 2.78, 4.11] &  -0.33 [-0.67, -0.17, 0.32] \\[0.5ex] 
430 & 93.4  & 99 & 56.5 & 98.1 & 100 & 0.52 & 		1.84 [1.91, 2.09, 2.3] & 1.09 [0.98, 1.16, 1.27]			\\[0.5ex]
353 & 88 & 72.6 &  88.8 &  90.8 & 98.8  &  0.51 & 	4.11 [1.71, 2.76, 4.02] & -0.08  [-0.42, 0.11, 0.56]						\\[0.5ex]
80 & 82.3 &  81.9 &  81.6 & 72.5&  87.6 &  0.35 & 5.73 [1.11, 2.75, 4.42] &  0.15 [-0.42, 0.48, 1.21]   \\[0.5ex]
461 & 99.6 &  33.4 & 99.1 & 98.4  & 100 & 0.32 & 3.8 [2.86, 4.16, 4.16] & 0.21 [-0.08, 0.03, 0.17]\\[0.5ex] 
379 & 93.8  & 99 &  98.4 &  91.9 &  36.8  & 0.28  & 	0.56 [0.55, 1.03, 1.3] & -0.75 [-1.42, -1.4, -0.36]					\\[0.5ex]
400 & 79.7  & 74.4 & 82.6 & 82.3 & 64.4 & 0.26 & 4.06 [0.78, 2.21, 4.54] & -1.41 [-1.54, 0.94, 1.54]  \\[0.5ex]  
356 & 93.8 & 95.7 &  93.4 & 26.1 & 100 & 0.20 & 3.79 [2.26, 3.45, 4.03] & -0.05 [-0.37, 0.04, 0.17]\\[0.5ex] 
176 & 95.3 & 98.7 & 98.2 & 6.7 &  100 & 0.06  &  1.18 [0.67, 0.92, 1.35] & 0.67 [0.14, 0.56, 1.02]					\\[0.5ex] 
105 & 75.8 & 19 & 96.5 & 37.7 & 99.7  & 0.06 & 2.56 [1.15, 1.71, 3.62] & -0.08 [0.12, 0.7, 0.97]\\[0.5ex] 
67 & 92.1 & 5.5 & 99.9 & 67.7 & 100 & 0.04 & 1.27 [0.85, 1.2, 1.64] & -1.31 [-1.5, -1.17, -0.86] \\[0.5ex] 
373 & 86.8 & 96.6 & 98.2 & 69.7 & 3 & 0.02 & 	0.49 [0.29, 0.56, 1.17] & -0.28  [-0.9, 0.12, 1.37]	\\[0.5ex] 
\hline 
\hline 
\end{tabular}
\end{footnotesize}
\end{center}
\end{table}

\begin{figure}[h!]
\begin{center}
\includegraphics[width=1\textwidth]{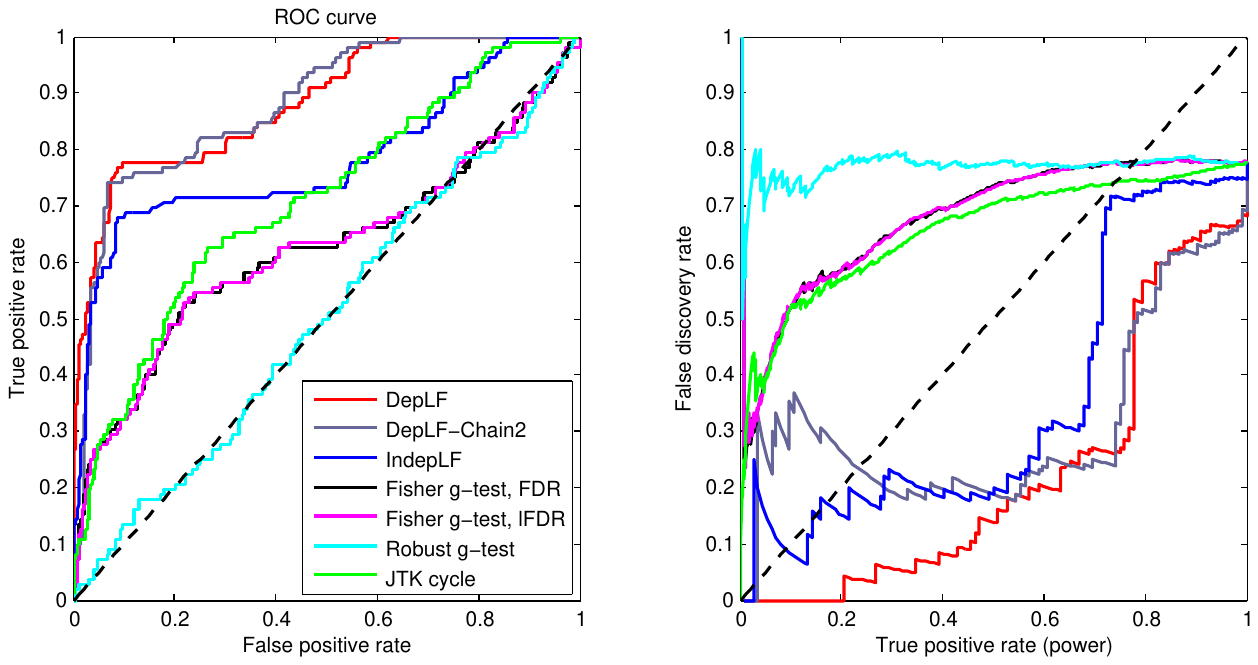}
 \end{center}
 \vspace{-6.5cm}
 \caption[ROC curve for identifying periodic signal in the simulated example with dependence across variables]{Left panel: ROC curve for identifying periodic signals in the simulated example with dependence across variables. Right panel: progression of false discovery rate (FDR) for different levels of true positive rate. Two chains initialized at over-dispersed starting values were run to assess the reproducibility of the results, and these chains correspond to the red and grey lines. The blue line corresponds to the ``independent'' version of our method as in (\ref{Independent_model}). Multiple testing for Fisher's $g$-test is done using tail area-based FDR and density-based local FDR (lFDR).}
\label{ROCdependence}
\end{figure}

To assess the performance of the proposed method, we compared our approach with Fisher's $g$-test \citep{Wichert2004}, robust $g$-test \citep{Olli2005}, and JTK cycle \citep{Hughes2010}. These methods test the hypothesis of ``absence of periodicity'' ($H_0$) versus ``signal is periodic'' ($H_1$) with unspecified period. Therefore, the comparison is made by evaluating the estimated probability that a protein is periodic (\ref{prob_periodic}). We also compared our method to its ``independent'' version that is,
	\begin{equation}\label{Independent_model}
	\by_i = \bB\btheta_i + \bC \bgamma_ i + \bepsilon_i, \quad \bepsilon_i \sim N(\bzr, \sigma_i^2\bI)
	\end{equation}
Model (\ref{Independent_model}) still accommodates local deviations and can detect periodicity, but does not accommodates dependence across variables by blocking inference on $\bLambda$ which is kept fixed to zero. Therefore, it is the default version of our model in scenarios of independence across variables. \\ 
\indent For every method, we ordered the p-values (or the estimated circadian probability for our approach) which show how strong the evidence is against the hypothesis of absence of pe\-ri\-odic signal. Based on this ordering, we picked-up the first $N_i$ variables from the ordered lists and compared the proportion the true positives, namely the proportion of periodic variables correctly identified as periodic, and the proportion of true negatives, namely the proportion of non-periodic variables correctly classified as non-periodic. Since predictions were periodic or non-periodic, a well-suited binary classification, we applied receiver operating characteristic (ROC) curves to compare the performances of the four algorithms by varying $N_i$ sequentially from $i = 1$ to $i = p$. The performance is measured by the area under the ROC curve criterion, with the larger area the better method (Figure~\ref{ROCdependence}). It is evident that our approach outperforms methods that do not directly accommodate dependence across variables. The grey ROC curve refers to a second chain for our model initialized at over-dispersed starting values; it is used to check the reproducibility of the results. The right panel of Figure~\ref{ROCdependence} shows the progression of the false discovery rate (FDR) as function of the true positive rate ({{power}}). The FDR is defined as the ratio between false positives, namely the number of variables falsely declared as periodic, and the number of positives, namely the total number of variables declared periodic. The spike at $0$ for Fisher's $g$-test and robust $g$-test means that the protein with smallest p-value, thus the first selected as periodic, is in fact a false positive. The more variables we include in the list of periodic variables, the more the power increases. A value of power equal to $1$ corresponds to detection of all truly periodic variables as periodic. For large values of power, our model achieves lower FDR than methods which do not accommodate dependence.\\
\indent Several chains were run to assess the sensitivity of the results to different choices of $a_\theta, a_\gamma, b_\theta, b_\gamma$, and other model parameters. In all cases, our approach achieved better performance than methods not accommodating dependence across variables.

\subsection{Independence across measurements} 
\label{Case2:independence}
\begin{figure}[h!]
\begin{center}
 \includegraphics[width=1\textwidth]{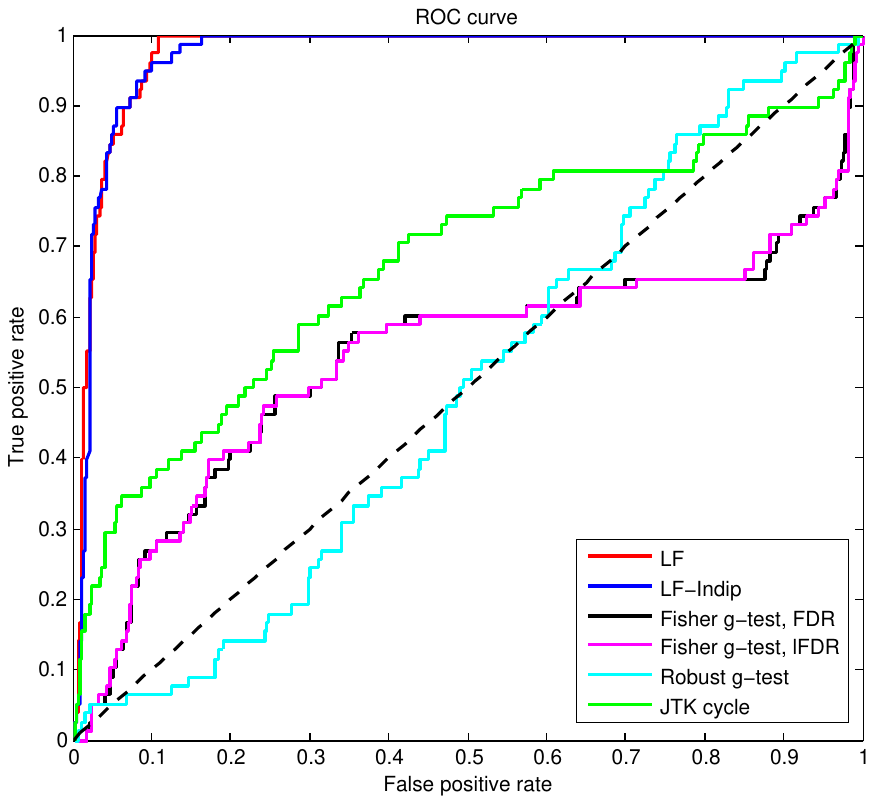}
 \end{center}
 \vspace{-5cm}
 \caption[ROC curve for identifying periodic signal in the simulated example with independence across variables]{ROC curve for identifying periodic signals in the simulated example with independence across variables. The blue curve refers to the ``independent'' version of our method obtained by keeping $\bLambda$ fixed to zero. Multiple testing for Fisher's $g$-test is done using tail area-based false discovery rate (FDR) and density-based local false discovery rate (lFDR).}
\label{ROCindependence}
\end{figure}

\begin{figure}[h!]
\begin{center}
 \includegraphics[width=1\textwidth]{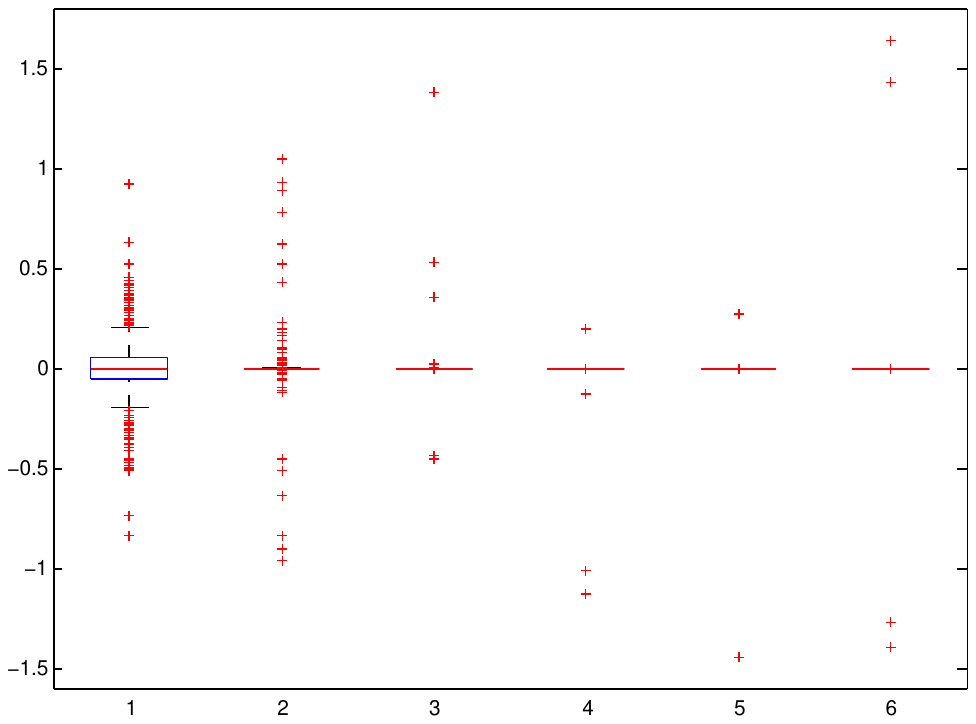}
 \end{center}
 \vspace{-6cm}
 \caption[Side-by-side boxplots of posterior mean estimates of the columns of the factor loading matrix $\bLambda$]{Side-by-side boxplots of posterior mean estimates of the columns of the factor loading matrix $\bLambda_{:, 1}$ (1), $\dots$, $\bLambda_{:, 6}$ (6), where $k = 6$ is the posterior mean estimate of the number of factors. The more these parameters are shrunk toward zero, the closer the model becomes to its ``independent'' version with $\bLambda \equiv \bzr$.}
\label{BoxplotLambda}
\end{figure}

For any modeling approach which accommodates dependence across variables, one concern is that if the true profiles are indeed independent, whether the ``unnecessarily sophisticated'' dependent-modeling approach can perform as good as the ``correct, independent'' model. \\
\indent To test the performance of our proposed method in such case, we simulated sample paths from the model but fixed $\bLambda \equiv \bzr$ such that at any time point $j = 1, \dots, T$ the variables were independent of each other, $\by^{(j)} \sim N(\bTheta\bbb_j+ \bGamma \bc_j, \bSigma)$, with $\bSigma$ diagonal. We increased $\sigma_i^2$ to $1, \forall i$ and generated the $p \times q$ true latent thresholds for $\bTheta$ independently from a Unif$(0,5)$. All remaining parameters were generated as described in Section \ref{Case1:dependence}. Of the $p = 500$ simulated trajectories, 4\% were  circadian and 78 were periodic with possible periods either $4, 6, 8, 12$ or $24$ hours. \\
\indent We run the Gibbs sampler described in Section \ref{Post} for 50000 iterations with a burn-in of 20000, and collected every 5th sample to thin the chain. The hyperparameters $a_\sigma$ and $b_\sigma$ for $\sigma_i^{-2}$ were 1 and 0.5, respectively, while $\rho = 3$, $a_1 = 2.1$, $a_2 = 3.1$, $a_\theta = a_\gamma = 1, \beta_\theta = \beta_\gamma = 5$ and used $k = 4$ as the starting number of factors. \\
\indent The additional complexity does not affect our method, which still outperforms Fisher's $g$-test, robust $g$-test and JTK cycle and performs at least as well as its corresponding independent version (Figure~\ref{ROCindependence}). The good performance has to be attributed to the shrinkage property of the MGPSP. Figure~\ref{BoxplotLambda} shows side-by-side boxplots of the posterior mean estimate of the factor loadings that is, the posterior means of $\{{\bLambda}_{1:p, 1}, {\bLambda}_{1:p, 2}, \dots, {\bLambda}_{1:p, k = 6}\}$, where $k = 6$ is the posterior mean of the estimated number of factors. Although this prior can not return exactly zero estimates for the components of $\bLambda$, the estimated factor loadings are small in magnitude, thus shrinking toward the truth (zero) the contribution of $\beeta\blambda_i$ in (\ref{yi_model}).

\section{Analysis of mouse liver mRNA data}
\label{protein}

\begin{figure}[h!]
\begin{center}
 \includegraphics[width = 16cm, height = 23cm]{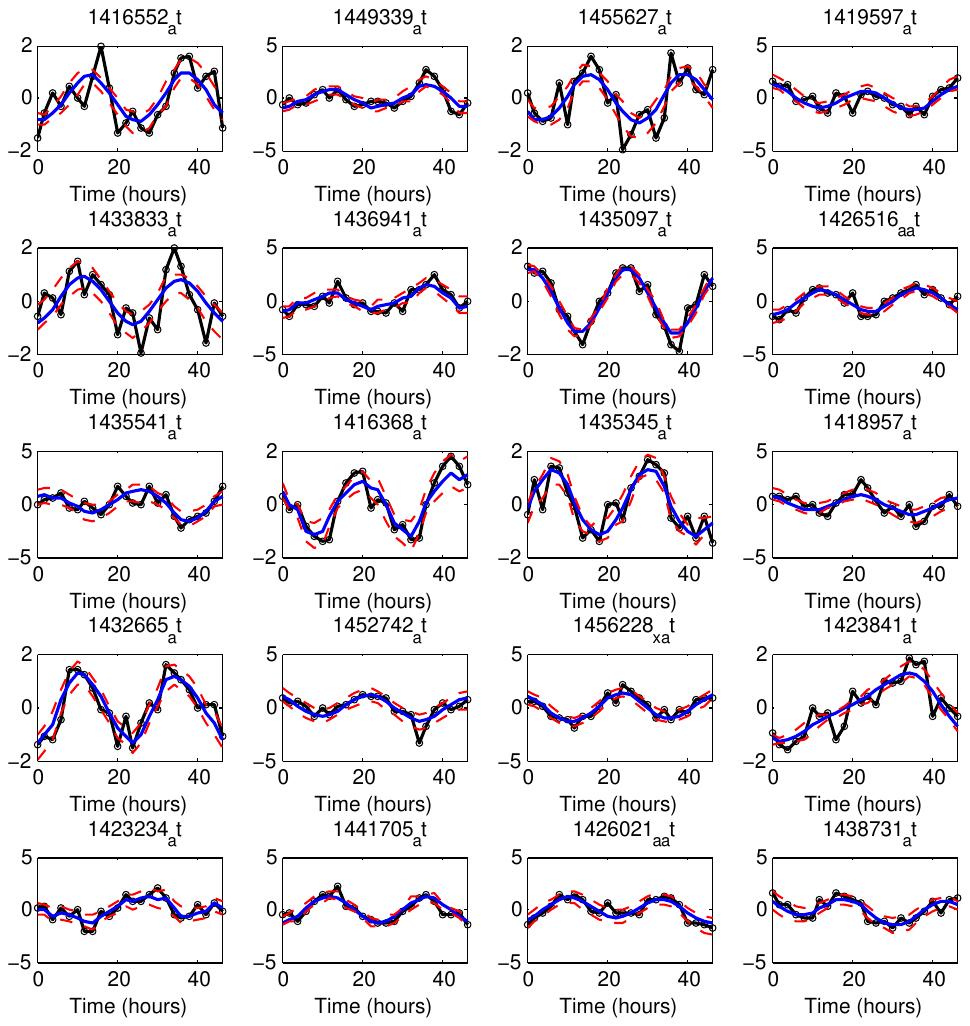}
 \end{center}
 \vspace{-4.8cm}
 \caption[The top 20 probes in the mouse liver microarray dataset according to the posterior probability of being periodic from Equation (\ref{prob_periodic})]{The top 20 probes in the mouse liver microarray dataset according to the posterior probability of being periodic from Equation (\ref{prob_periodic}). The black trajectory connects the true expression levels, the blue line represents the posterior mean, and red lines are the pointwise 95\% credible intervals. The $y$-axis denotes the $\log$-transformed and normalized expression levels.}
\label{top20periodic}
\end{figure}

\begin{figure}[h!]
\begin{center}
 \includegraphics[width=.70\textwidth]{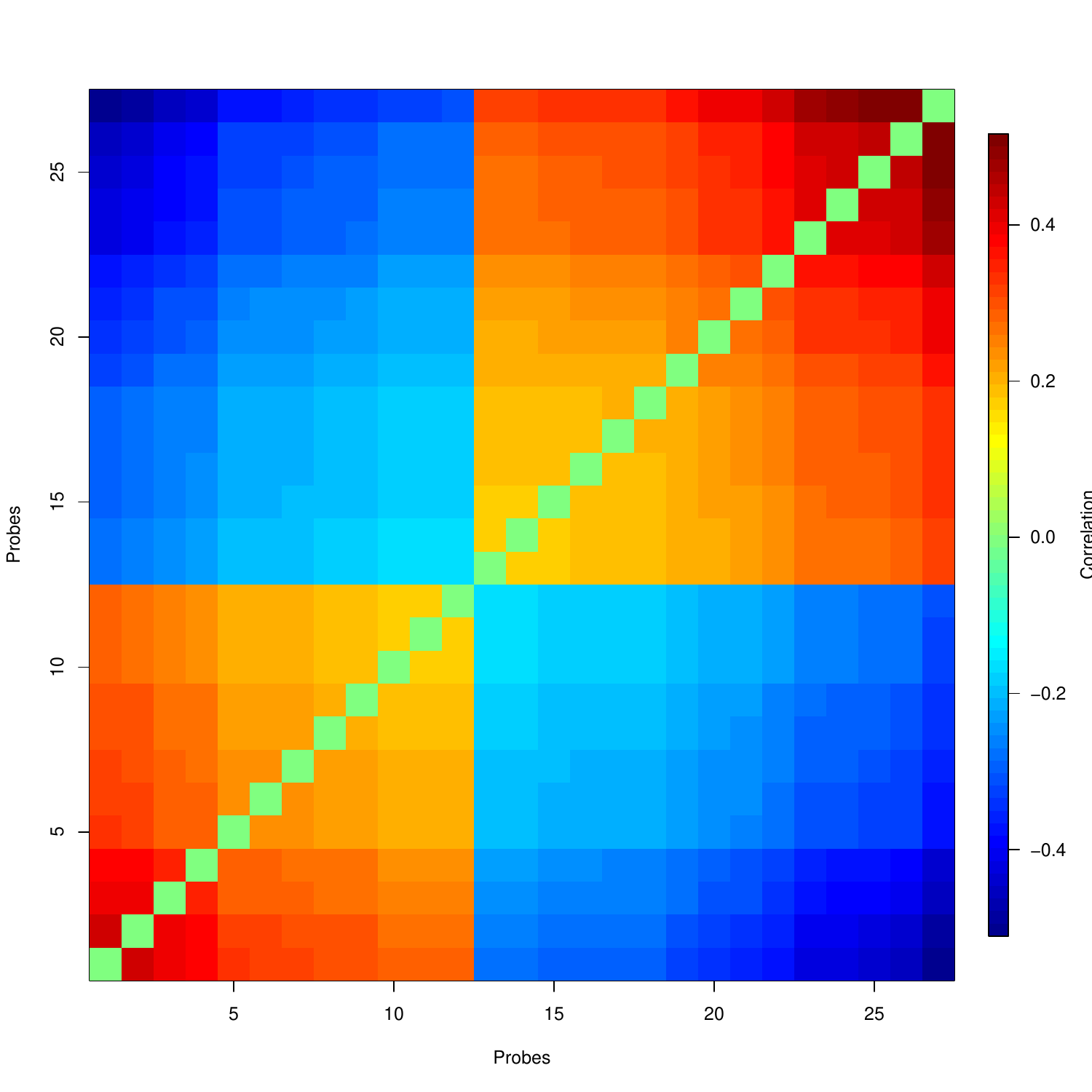}
  \end{center}
\vspace{-0.3cm}
 \caption[Correlation structure of the 27 proteins with strongest inferred correlation (of or above $0.30$)]{Correlation structure of the 27 proteins with strongest inferred correlation (of or above $0.30$). To improve visibility, the diagonal of the correlation matrix was set to 0.}
\label{corrstructure}
\end{figure}

\begin{figure}[h!]
\begin{center}
 \includegraphics[width=.8\textwidth]{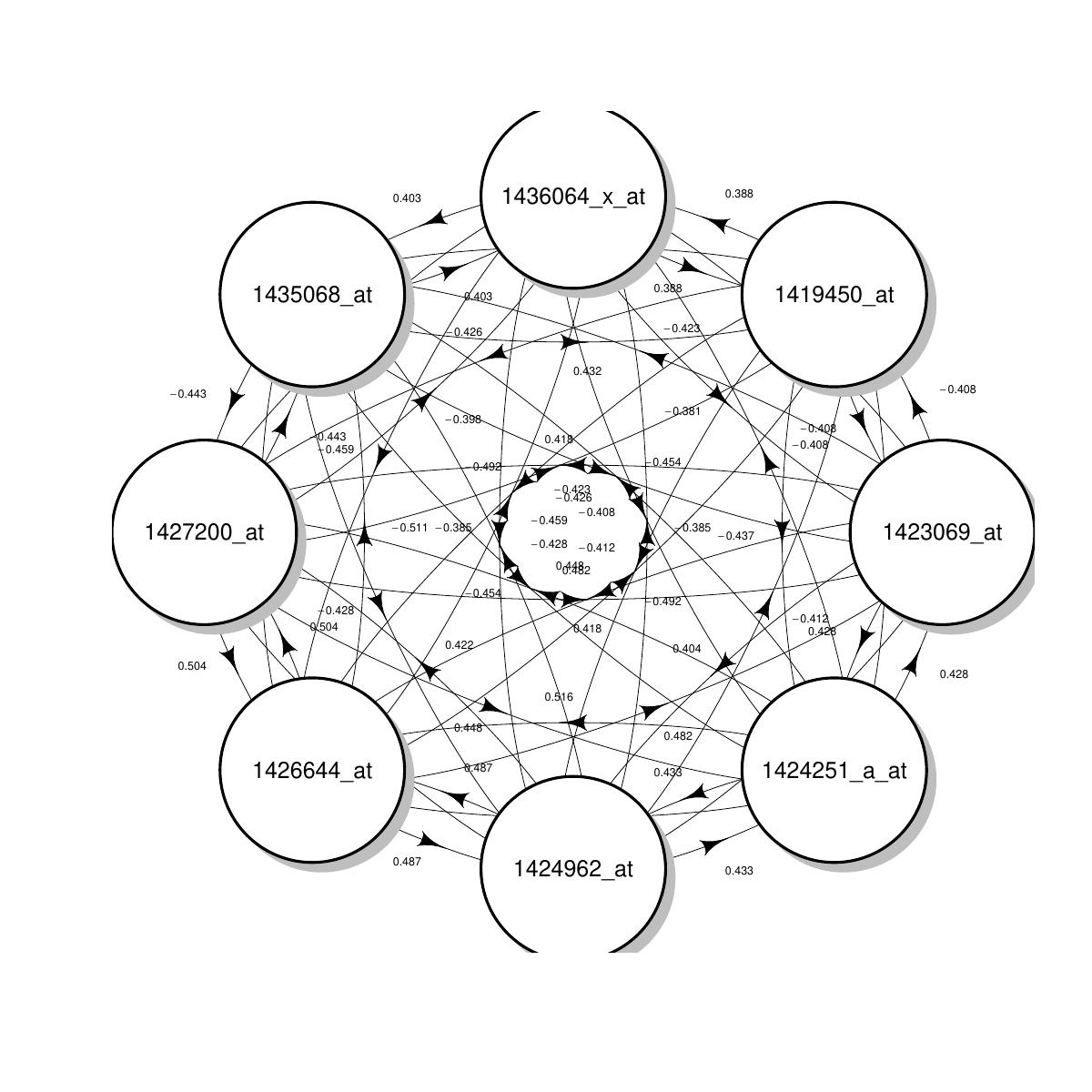}
  \end{center}
\vspace{-2cm}
 \caption[Network graph of probes with estimated correlation of or above $0.45$ in absolute value]{Network graph of probes with estimated correlation of or above $0.45$ in absolute value.}
\label{network}
\end{figure}

\begin{figure}[t!]
\begin{center}
 \includegraphics[width = 1.05\textwidth]{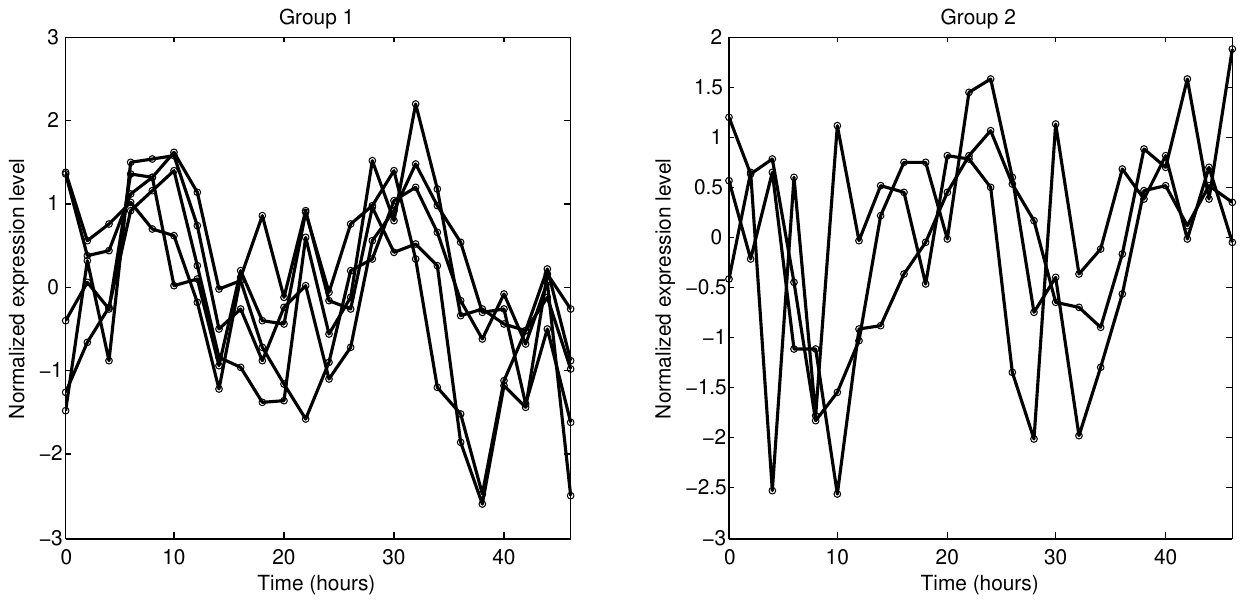}
  \end{center}
\vspace{-7.5cm}
 \caption[True trajectories of proteins with largest inferred correlations. ]{True trajectories of proteins with largest inferred correlations. These trajectories were obtained by linearly connecting the observed normalized expression levels. Group 1 includes probes 1423069 AT, 1424251 A AT, 1424962 AT, 1426644 AT, 1427200 AT; group 2 includes probes 1419450 AT, 1435068 AT, 1436064 X AT. Proteins within each group are positively correlated with each other, whereas correlation is negative across groups.}
\label{truetrajcorr}
\end{figure}

We apply our method to a real dataset generated from the work of \cite{Fred2013}. The goal of the study was to assess whether the circadian clock could coordinate the transcription of messenger RNA (mRNA) in mouse liver. In the experiment, C57B1/6J male mice between 10 and 12 weeks of age were used. Mice were maintained under standard animal housing conditions, with free access to food and water and in 12 hours light/12 hours dark cycles. However, mice were fed only at night during 4 days before the experiment to reduce the effects of feeding on rhythm. In the case of rodents, it is in fact during the night period that animals are active and consume food. Liver polysomal and total RNAs were extracted independently from two mice sacrificed every 2 hours during 48 hours. 3 $\mu$g of polysomal and total RNAs from each animal from each time point were pooled. The 6 $\mu$g of polysomal and total mRNAs were used for the synthesis of biotinylated complimentary RNAs (cRNAs) according to Affymetrix protocol, and the fluorescence signal was analyzed with Affimetrix software (refer to \cite{Fred2013} for more details on the study). Data are deposited on the Gene Expression Omnibus database under the reference GSE33726. In the original study, the rhythmic characteristics of the expression of each gene or protein were assessed by a Cosinor analysis \citep{Nelson1979}, and a rhythm was detected if the null hypothesis was rejected with $p$-value $< 0.05$. A period of 24 hours was considered a priori. Consequently, the authors validated as circadian a subset of the detected genes and encoded proteins in laboratory. The authors concluded that the circadian clock influences the temporal translation of a subset of mRNAs mainly involved in ribosome biogenesis. In addition, the circadian clock appeared to regulate the transcription of ribosomal protein mRNAs and ribosomal RNAs. \\
\indent We considered a randomly selected subset of $p = 1000$ proteins from the full dataset. The raw expression levels were $\log$-transformed and normalized to zero-mean following standard practice. We run the Gibbs sampler described in Section \ref{Post} for 50000 iterations with a burn-in of 20000, and collected every 5th sample to thin the chain. The hyperparameters $a_\sigma$ and $b_\sigma$ for $\sigma_i^{-2}$ were 1 and 0.5, respectively, while $\tilde{T} = 20, \rho = 3$, $a_1 = 2.1$, $a_2 = 3.1$, $a_\theta = a_\gamma = 1, \beta_\theta = 6, \beta_\gamma = 10$ and used $k = 8$ as the starting number of factors. \\
\indent Using our model, we ranked the probes by their posterior probability of being periodic (\ref{prob_periodic}), and Figure~\ref{top20periodic} shows the top 20 probes. Of the top 20 (50) probes, 19 (42) of them also rank among the top 20 (50) by posterior probability of being circadian. We also tested the probes with Fisher's $g$-test, JTK cycle, and the independent version of our approach and compared the top 100 probes by evidence against the null hypothesis of absence of periodicity (by p-value or posterior probability of being periodic). Of these top 100 probes, 50 are shared with JTK cycle, 47 with Fisher's $g$-test, and 61 with the independent version of our method. All together, the four methods agreed on a common set of 36 probes as most likely to be periodic. 
 \indent A key attractive feature of our model is the accommodation of dependence across proteins, thus it becomes of interest to examine the inferred correlation structure of the 1000 proteins from the estimated covariance $\bOmega = \bLambda \bLambda^\top + \bSigma$.  Most of the estimated correlations are small in magnitude and only 104 pairs of probes have correlation equal or above $0.30$ (in absolute value). These ``major'' correlations are controlled by only 27 (dominant) probes linked (positively or negatively) to each other as shown in Figure~\ref{corrstructure}. By inspecting the plot, one can envision two groups of probes: one group with proteins $1-13$ in the bottom left corner and the second group with proteins $14-27$ in the top right corner. Correlation is positive within each group and negative across groups. To be more conservative, we can further restrict to a set of 8 proteins with estimated correlation of or above $0.45$ (in absolute value) as shown in Figure~\ref{network}. Again, these 8 proteins divide into 2 groups: group 1 with probes 1423069 AT, 1424251 A AT, 1424962 AT, 1426644 AT, 1427200 AT; and group 2 with probes 1419450 AT, 1435068 AT, 1436064 X AT. Proteins within each group are positively correlated with each other, whereas the correlation is negative across groups. Proteins in each group exhibit a common pattern in normalized expression level (Figure~\ref{truetrajcorr}). Proteins in group 1 exhibit a dip in their normalized expression levels between 10 and 30 hours, as opposed to the peak that probes in group 2 exhibit. Although local deviation seem to emerge in the trajectories of proteins in group 2, the sequence of dips and peaks over time seems overall reversed in the two groups.
\section{Conclusions}
A flexible Bayesian methodology for periodicity detection has been developed and applied to large-scale circadian gene expression studies. It employs a Fourier basis expansion with variable selection priors on the basis coefficients to model the time course gene expression trajectories and identify rhythmic genes. The key statistical contribution is to accommodate the potential dependence in the trajectories in terms of latent factors. Our construction allows to infer groups of co-expressed collections of genes and verify relationships within and across groups. Further, accommodating dependence helps identifying weaker patterns appearing is expression profiles by sharing information across genes. Simulation studies show that our construction gives significantly improved performance over widely-used rhythmicity detection techniques that do not directly accommodate for dependence across genes.

\bibliographystyle{agsm}
 \bibliography{circadian_literature}

\end{document}